\documentclass[journal]{IEEEtran}

\usepackage{cite, url, amsmath, amssymb, array, balance, times}
\usepackage{algorithm, algorithmic, calc, color, multicol}
\usepackage{graphics, epsfig, epsf, psfrag, subfigure, graphicx}
\usepackage{amsfonts, color}
\usepackage{textcomp}
\usepackage{mathrsfs}
\usepackage{stmaryrd}
\usepackage{setspace, verbatim}

\DeclareGraphicsExtensions{.eps, .pdf, .png, .jpg}   

\newtheorem{example}{Example}

\newcommand{\Jset}{\mathcal{J}}
\newcommand{\Sset}{\mathcal{S}}
\newcommand{\Kset}{\mathcal{K}}
\newcommand{\Aset}{\mathcal{A}}
\newcommand{\Pset}{\mathcal{P}}
\newcommand{\Cset}{\mathcal{C}}
\newcommand{\Nset}{\mathcal{N}}
\newcommand{\Hset}{\mathcal{H}}
\newcommand{\Iset}{\mathcal{I}}

\newcommand{\Zset}{\mathcal{Z}}
\newcommand{\Qset}{\mathcal{Q}}

\begin{document}


\title{Network Coding-Aware Queue Management\\
for TCP Flows over Coded Wireless Networks}

\author{
\IEEEauthorblockN{Hulya Seferoglu, Athina Markopoulou\\}
\IEEEauthorblockA{EECS Dept, University of California, Irvine\\
{\tt \{hseferog, athina\}@uci.edu}}
}

\maketitle

\begin{abstract}
We are interested in unicast traffic over wireless networks that employ constructive inter-session network coding, including single-hop and multi-hop schemes. In this setting, TCP flows do not fully exploit the network coding opportunities due to their bursty behavior and due to the fact that TCP is agnostic to the underlying network coding. In order to improve the performance of TCP flows over coded wireless networks, we take the following steps. First, we formulate the problem as network utility maximization and we present a distributed solution. Second, mimicking the structure of the optimal solution, we propose a "network-coding aware" queue management scheme (NCAQM) at intermediate nodes; we make no changes to TCP or to the MAC protocol (802.11). We demonstrate, via simulation, that NCAQM significantly improves TCP performance compared to TCP over baseline schemes. 
\end{abstract}


\vspace{-10pt}
\section{Introduction}
Wireless environments naturally lend themselves to network coding, thanks to the inherent broadcast and overhearing capabilities of the wireless medium. We are particularly interested in wireless mesh networks that employ constructive network coding schemes (such as COPE \cite{cope} and BFLY \cite{BFLY}), to mention some concrete examples). We consider unicast flows (particularly TCP, which is the dominant traffic type today) transmitted over such coded wireless networks.

In this setting, it has been demonstrated that network coding can significantly increase throughput \cite{chou-unicast, cope}. However, it has also been observed \cite{cope} that TCP does not exploit the full potential of the underlying network coding, mainly due to its bursty behavior. Rate mismatch between flows can significantly reduce the coding opportunities, as there may not be enough packets from different flows at intermediate nodes to code together. One possible solution is to artificially delay packets at intermediate nodes \cite{towsley_secon_2008}, until more packets arrive and can be coded together. However, the throughput increases with small delay (due to more coding opportunities), but decreases with large delay (which reduces the TCP rate); the optimal delay depends on the network topology and the background traffic and also may change over time. Thus, in many practical networking scenarios, introducing delay at intermediate nodes is not practical.

We consider the same problem but we propose a different approach. Our main observation is that the mismatch between flow rates is due to the dynamic/bursty nature of TCP. Therefore, the problem can be eliminated by making modifications to congestion control mechanisms (at the end-points) and/or to queue management schemes (at intermediate nodes) to make them network coding-aware (in the sense that they can match the rates of flows coded together). Based on this observation, we take the following steps.

First, we formulate congestion control for unicast flows over wireless networks with inter-session network coding within the network utility maximization (NUM) framework \cite{tutorial_doyle, book_srikant}. We assume that a known constructive network coding scheme is deployed in a wireless mesh network; examples include COPE \cite{cope} for one-hop network coding and BFLY \cite{BFLY} for two-hop network coding. The optimal solution of the NUM problem decomposes into several parts, each of which has an intuitive interpretation, such as rate control, queue management, and scheduling.

Second, motivated by the analysis, we propose modifications to congestion control mechanisms, so as to mimic the optimal solution of the NUM problem and to fully exploit the potential of network coding. It turns out that the optimal solution dictates minimal and intuitive implementation changes. We propose a network coding-aware queue management scheme at intermediate nodes (NCAQM), which stores coded packets and drops packets based on both congestion state and network coding. We note that the queues at intermediate nodes, which are already used for network coding, are a natural place to implement such changes with minimal implementation cost. In contrast, we do not propose any practical modifications to TCP or MAC (802.11) protocols,
which significantly simplifies practical deployment of our proposal. Finally, we evaluate our proposal via simulation in GloMoSim \cite{glomosim} and we show that TCP over NCAQM significantly outperforms TCP over baseline schemes ({\em e.g.,} doubles the throughput improvement in some scenarios), 
and achieves near-optimal performance.

The rest of the paper is organized as follows. Section~\ref{sec:related} discusses related work. Sections \ref{sec:system}-\ref{sec:performance} focus on wireless networks with one-hop network coding:
Section~\ref{sec:system} presents the system model; Section~\ref{sec:opt2} presents the optimization problem and solution; Section~\ref{sec:algs} presents the design of our network coding-aware queue management scheme (NCAQM); Section~\ref{sec:performance} presents simulation results. Section~\ref{sec:opt_multi_hop} extends our framework to multi-hop network coding. Section~\ref{sec:conclusion} concludes the paper. Appendix A presents numerical results for the convergence of the solution.

\vspace{-5pt}
\section{\label{sec:related}Related Work}
\vspace{-5pt}
This paper  builds on top of constructive network coding schemes in wireless mesh networks. We rely on such a given scheme to provide the available coded and uncoded flows to higher layers. We then seek to optimize the treatment of these flows at the end-points and/or at intermediate nodes so as to maximize network coding opportunities.

{\em COPE and follow-up work.}
COPE \cite{cope} is a constructive network coding scheme
for one-hop network coding across unicast sessions. Our framework can also consider any other constructive scheme for inter-session NC, such as BFLY \cite{BFLY} or tiling approaches \cite{tiling}. COPE has generated a lot of interest in the research community. Some researchers tried to model and analyze COPE \cite{proutiere_scheduling}, \cite{how_many_packets_infocom_2008}, \cite{sudipta_sengupta_infocom_2007}. Some others proposed new coded wireless systems, based on the idea of COPE \cite{practical_nc_dong}, \cite{BFLY}. Zhao and Medard tried to explain and improve COPE's performance by looking at its interaction with MAC fairness \cite{medard_ita}.
We note that the authors of COPE had noticed the problem with TCP performance over COPE. As discussed in the introduction, \cite{towsley_secon_2008} addressed the problem of rate mismatch between flows that are coded together, by delaying packets. Here, we take a different approach and we create coding opportunities via queue management and congestion control.
More specifically, we aim at improving TCP performance over COPE by complementing it with a network coding-aware queue management scheme (NCAQM).

{\em NUM in coded systems.} Our analysis falls within the classic framework of network utility maximization (NUM) \cite{tutorial_doyle}.  A significant body of work has looked at the joint optimization of intra- or inter-session NC of unicast flows. For example,  in \cite{minimum_cost_multicast}, minimum cost multicast over network coded wireline and wireless networks was studied. This work was extended for rate control in \cite{opt_multicast_nc} for wireline networks. The rate region of multicast flows when network coding is used is studied in \cite{opt_framework_in_wireless, opt_models_in_wireless}.
The most closely related to this paper are resource allocation problems for unicast flows. For example, rate control, routing, and scheduling for generation-based intra-session network coding over wireless networks is considered in \cite{bozidar_infocom_2008}. Optimal scheduling and optimal routing for COPE are considered in \cite{proutiere_scheduling} and \cite{sudipta_sengupta_infocom_2007}, respectively. Network utility maximization is used in \cite{pairwise} for end-to-end pairwise inter-session network coding. Energy efficient opportunistic inter-session network coding over wireless are proposed in \cite{energy_ho}, following a node-based NUM formulation and its solution based on back-pressure. A linear optimization framework for packing butterflies is proposed in \cite{poison_antidote}.

Compared to prior NUM problems in coded networks, we focus on the congestion control problem for multiple unicast flows over wireless with a given inter-session network coding scheme. The most similar formulation is  \cite{opt_multicast_nc}, but for intra-session network coding.

{\em Protocol design.} To the best of our knowledge, our work is the first, to take the step from theory (optimization) to practice (protocol design), specifically for the problem of congestion control over inter-session network coding. We propose implementation changes, which have a number of desired features: they are justified and motivated by analysis, they perform well (double the throughput in simulations), and they are minimal (only queue management is affected, while TCP and MAC remain intact).

{\em Comparison to our prior work.} This paper is an improved and  extended version of our conference paper that was presented in NetCod 2010 \cite{hs_netcod}. It includes significantly extended sections on simulations of  performance (Sections VI and VII) as well as new numerical results on the convergence (Appendix A) of our schemes. It also extends the framework from one-hop to multi-hop network coding (Section VII).

In another piece of recent work \cite{hs_intra_inter}, we studied a related but orthogonal aspect: we added intra-session redundancy to inter-session network coding in order to deal with wireless losses and to eliminate the need to know the state of the neighbors. In contrast, in this paper we consider only inter-session coding and we focus on the interaction between local (coding and queue management) and end-to-end (TCP) schemes, which was out of the scope of \cite{hs_intra_inter}.

\vspace{-10pt}
\section{\label{sec:system}System Model}
\textbf{Sources/Flows.} Let $\Sset$ be the set of unicast flows between some source-destination pairs. Each flow $s \in \Sset$ is associated with a rate $x_{s}$ and a utility function $U_{s}(x_{s})$, which we assume to be a strictly concave function of $x_{s}$. The goal is to maximize the total utility function $U_{t} = \sum_{s \in \Sset} U_{s}(x_{s})$.

\textbf{Wireless Network.}
A hyperarc $(i,\Jset)$ is a collection of links from node $i \in \Nset$ to a non-empty set of next-hop nodes $\Jset \subseteq \Nset$ that are interested in receiving the same network code through a broadcast transmission from $i$. A hypergraph $\Hset=(\Nset,\Aset)$ represents a wireless mesh network, where $\Nset$ is the set of nodes and $\Aset$ is the set of hyperarcs. For simplicity, $h=(i,\Jset)$ denotes a hyperarc, $h(i)$ denotes node $i$ and $h(\Jset)$ denotes node $\Jset$, {\em i.e.}, $h(i)=i$ and $h(\Jset) = \Jset$. We use these terms interchangeably in the rest of the paper.

Due to the shared nature of the wireless media, transmission over different hyperarcs may interfere with each other. We consider the protocol model of interference \cite{gupta_interference_model}, according to which, each node can either transmit or receive at the same time and all transmissions in the range of the receiver are considered as interfering. Given a hypergraph $\Hset$, we can construct the conflict graph $\Cset = (\Aset, \Iset)$, whose vertices are the hyperarcs of $\Hset$ and edges indicate interference between hyperarcs. A clique $\Cset_{q} \subseteq \Aset$ consists of several hyperarcs, at most one of which can transmit at the same time without interference.

\textbf{Network Coding:}
We assume that intermediate nodes use COPE \cite{cope} for one-hop opportunistic network coding\footnote{Note that we present the multi-hop extension in Section~\ref{sec:opt_multi_hop}.}. Each node $i$ listens all transmissions in its neighborhood, stores the overheard packets in its decoding buffer, and periodically advertises the content of its decoding buffer to its neighbors. Then, when a node $i$ wants to transmit a packet, it checks or estimates the contents of the decoding buffer of its neighbors. If there is a network coding opportunity, the node combines the relevant packets using simple coding operations (XOR) and broadcasts the combination to $\Jset$. Note that it is possible to construct more than one network code over a hyperarc $(i,\Jset)$. Let $\Kset_{i,\Jset}$ be the set of network codes over a hyperarc $(i,\Jset)$. Let $\Sset_{k} \subseteq \Sset$ be the set of flows, whose packets are coded together using code $k \in \Kset_{i, \Jset}$ and broadcast over $(i,\Jset)$.

\textbf{Routing:}
We consider that each flow $s \in \Sset$ follows a single path $\Pset_{s} \subseteq \Nset$ from the source to the destination. This path is pre-determined by a routing protocol, {\em e.g.}, OLSR or AODV, and given as input to our problem. However, note that several different hyperarcs may connect two consecutive nodes along the path. We set an indicator function $H_{i,\Jset}^{s,k}=1$  if flow $s$ is transmitted through hyperarc $(i,\Jset)$ using network code $k \in \Kset_{i,\Jset}$. Otherwise, $H_{i,\Jset}^{s,k} = 0$.

\begin{figure}
\centering \vspace{-5pt}
\includegraphics[width=7cm]{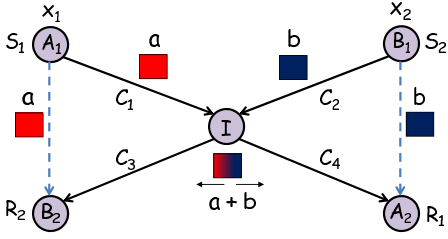}
\vspace{-5pt}
\caption{``X topology''. Source $S_1$ transmits a flow with rate $x_1$ to receiver $R_1$ and source $S_2$ transmits a flow with rate $x_2$ to receiver $R_2$, over the intermediate node $I$. $A_1$ and $B_1$  transmit their packets $a$ and $b$, in two time slots, and node $I$ receives them. Furthermore, $A_2$ overhears $b$ and $B_2$ overhears $a$, because  $A_1-B_2$ and $B_1-A_2$ are in the same transmission range and they can overhear each other. In the next time slot, $I$ broadcasts the network coded packet, $a \oplus b$ over hyperarc $(I,\{A_2,B_2\})$. Since $A_2$ and $B_2$ have overheard $b$ and $a$, they can decode their packets $a$ and $b$, respectively.
}
\label{fig:x_topology}
\vspace{-15pt}
\end{figure}

\begin{example}\label{ex1}
The example shown in Fig.~\ref{fig:x_topology} illustrates the problem we consider. Since $I$ can transmit $a \oplus b$ in one time slot, instead of $a, b$ in two time-slots, network coding has the potential to improve throughput. However, if there is mismatch between the rates $x_1, x_2$ of the two flows, $I$ may not have packets from the two flows to code together at all times, and thus does not exploit the full potential of network coding. We confirmed this intuition through simulations in this example topology. When the buffer size was set to $10$ packets at each node and the bandwidth was $1Mbps$ for each link, we observed that $50\%$ of the time, there were no packets from the two flows at the same time at node $I$ to code together. For smaller queue sizes and larger transmission rates, there were even fewer coding opportunities. This means that there is potential for improvement by updating the protocols so as to mitigate the rate mismatch between TCP flows. This is the observation that motivates this paper.
\hfill $\Box$
\end{example}

\vspace{-5pt}
\section{\label{sec:opt2}Optimal Congestion Control}
\subsection{Problem Formulation}
The objective is to maximize the total utility function, by appropriately selecting: the flow rates $x_s$ at sources $s \in \Sset$; their traffic splitting parameter $\alpha_{h}^{s,k}$ (following the terminology of \cite{opt_multicast_nc}) into network codes $k \in \Kset_{h}$ over hyperarc $h$ at intermediate nodes; and the percentage of time $\tau_{h}$ each hyperarc is used:

\begin{align} \label{opt:eq1}
\max_{\boldsymbol x, \boldsymbol \alpha, \boldsymbol \tau} & \sum_{s \in \Sset} U_{s}(x_{s}) \nonumber \\
\mbox{s.t.}  & \sum _{k \in \Kset_{h}} \max_{s \in \Sset_{k}} \{ H_{h}^{s,k} \alpha_{h}^{s,k} x_{s}\} \leq R_{h} \tau_{h}, \mbox{   } \forall  h  \in \Aset \nonumber \\
       & \sum_{h(\Jset) | h \in {\Aset}} \sum_{k \in \Kset_{h} \mid s \in \Sset_{k}} \alpha_{h}^{s,k}  = 1, \mbox{   }  \forall s \in \Sset, i \in \Pset_s \nonumber \\
       & \sum_{h \in \Cset_q} \tau _{h} \leq \tau,  \mbox{   }  \forall \Cset_q \subseteq \Aset
\end{align}
The first constraint is the capacity constraint. $H_{h}^{s,k} \alpha_{h}^{s,k} x_{s}$  indicates the part of flow rate $x_s$ allocated to the $k$-th network code over hyperarc $h$. The rate of the $k$-th network code is the maximum rate among flows $s \in \Sset_{k}$  coded together in code $k$: $\max_{s \in \Sset_{k}} \{ H_{h}^{s,k} \alpha_{h}^{s,k} x_{s}\}$ \cite{minimum_cost_multicast}. Different network codes $k \in \Kset_{h}$ over $h$ share the available capacity $R_{h}\tau_{h}$, where $R_{h}$ is the transmission capacity of $h$; since $h$ is a set of links, $R_{h}$ is the minimum: $R_{h} = \min_{j \in h(\Jset)} \{R_{i,j}\xi_{i,j}\}$ where $R_{i,j}$ is the capacity of link $(i,j)$, and $\xi_{i,j}$ is the probability of successful transmission over link $(i,j)$. The second constraint is the flow conservation constraint: at every node $i$ on the path $\Pset_s$ of source $s$, the sum of $\alpha_{h}^{s,k}$ over all network codes and hyperarcs should be equal to $1$. Indeed, when a flow enters a particular node $i$, it can be transmitted to its next hop $j$ as part of different network coded and uncoded flows. The third constraint is due to interference. As mentioned, $\tau_{h}$ is the percentage of time $h$ is used. Its sum over all hyperarcs in a clique should be less than an over-provisioning factor, $\gamma \le 1$, because all hypearcs in a clique interferes, and should time share the medium.

\subsection{Solution}
By relaxing the capacity constraint in Eq.~(\ref{opt:eq1}), we get the Lagrangian:
\begin{align} \label{opt:eq1_Lagrange1}
& L(\boldsymbol x, \boldsymbol \alpha, \boldsymbol \tau, \boldsymbol q)= \nonumber \\
& \sum_{s \in \Sset} U_{s}(x_{s}) - \sum _{h \in \Aset} q_{h} \left( \sum _{k \in \Kset_{h}} \max_{s \in \Sset_{k}} \{ H_{h}^{s,k} \alpha_{h}^{s,k} x_{s}\} - R_{h} \tau_{h} \right)
\end{align}
 where $q_{h}$ is the Lagrange multiplier, which can be interpreted as the queue size at hyperarc $h$, as discussed later.
To decompose the Lagrangian, we rewrite $\max_{s \in \Sset_{k}} \{ H_{h}^{s,k} \alpha_{h}^{s,k} x_{s}\}$  as $\max_{m_{h}^{s,k}} { \sum_{s \in \Sset_{k}} H_{h}^{s,k} \alpha_{h}^{s,k} x_{s} m_{h}^{s,k}}$ {\em s.t.} $\sum_{s \in \Sset_{k}} m_{h}^{s,k} = 1$, where $m_{h}^{s,k}$ is a new variable, which we call the the {\em dominance indicator}. It indicates whether the source $s$ has the maximum rate among all flows coded together in the $k$-th network code, or not. In the next section, we will see that only the dominant flow in a network code needs to back-off during congestion.

The Lagrange function in Eq.~(\ref{opt:eq1_Lagrange1}) is not strictly concave in $m_{h}^{s,k}$ and this causes oscillation in its solution.
We use the proximal method \cite{bertsekas_parallel_dist_comp_book} to eliminate oscillations;
\begin{align} \label{opt:intermediate_problem}
\max_{\boldsymbol m} & \sum_{s \in \Sset_{k}} (H_{h}^{s,k} \alpha_{h}^{s,k} x_s m_{h}^{s,k} - c (m_{h}^{s,k} - \mu_{h}^{s,k})^2)   \nonumber \\
\mbox{s.t.}  & \sum_{s \in \Sset_{k}} m_{h}^{s,k} = 1,
\end{align} where $c$ is a constant and $\mu_{h}^{s,k}$ is an artificial variable of the proximal method \cite{bertsekas_parallel_dist_comp_book}. Its value is set to $m_{h}^{s,k}$ periodically. Let $(m_{h}^{s,k})^{*}$ be the solution to this problem.

By rewriting the summation $\sum_{k \in \Kset_{h}}$ $\sum_{s \in S_{k}}$ as $\sum_{s \in \Sset}$ $\sum_{k \in K_{h} \mid s \in \Sset_{k}}$, the Lagrange function in Eq.~(\ref{opt:eq1_Lagrange1}) can be expressed as:
$ L(\boldsymbol x, \boldsymbol \alpha, \boldsymbol \tau, \boldsymbol q) =
 \sum_{h \in \Aset} q_{h} R_{h} \tau_{h} + \sum_{s \in \Sset} \left(U_{s}(x_{s}) - x_s \sum_{h \in \Aset} \sum_{k \in \Kset_{h} \mid s \in \Sset_{k}} q_{h} H_{h}^{s,k} \alpha_{h}^{s,k} (m_{h}^{s,k})^{*}\right)$.
Now, we can decompose the Lagrangian into the following intuitive problems: rate control, traffic splitting, scheduling, and parameter update (queue management).

\textbf{Rate Control.} First, we solve the Lagrangian w.r.t $x_s$:
\begin{equation} \label{opt:eq1_rateControl1}
x_s = ({U'_{s}})^{-1} \left( \sum_{h \in \Aset} \sum_{k \in \Kset_{h} \mid s \in \Sset_{k}}  q_{h} H_{h}^{s,k} \alpha_{h}^{s,k} (m_{h}^{s,k})^{*} \right),
\end{equation}
where $({U'_{s}})^{-1}$ is the inverse function of the derivative of $U_{s}$. If we define $w_{h}^{s} = \sum_{k \in \Kset_{h} | s \in \Sset_{k}}$  $H_{h}^{s,k}\alpha_{h}^{s,k}(m_{h}^{s,k})^{*}$ and $q_{h(i)}^{s} = \sum_{h(\Jset) | h \in \Aset}q_{h}w_{h}^{s}$, the rate $x_s$ can be expressed as $x_s = ({U'_{s}})^{-1} ( \sum_{i \in \Pset_{s}}  q_{i}^{s} )$, noting that $i = h(i)$.

In the special case where proportional fairness is desired, $U_s(x_s) = \log (x_s), \forall s \in \Sset$, leading to $x_s = \left(\sum_{i \in \Pset_{s}} q_{i}^{s}\right)^{-1}$, {\em i.e.,} $x_s$ is inversely proportional to the total network coded queue sizes over the path of flows $s$, which we will be explained later.

\textbf{Traffic Splitting.} Second, we solve the Lagrangian for $\alpha_{h}^{s,k}$:
at each node $i$ along the path ({\em i.e.}, $i \in \Pset_{s}$), the traffic splitting problem can be expressed as
\begin{align} \label{opt:eq1_trafficSplit}
\min_{\boldsymbol \alpha} & \sum_{h(J)|h \in \Aset} \sum_{k \in K_{h} | s \in \Sset_{k}} q_{h} H_{h}^{s,k} (m_{h}^{s,k})^{*} \alpha_{h}^{s,k} \nonumber \\
\mbox{s.t. } & \sum_{h(J)|h \in \Aset} \sum_{k \in K_{h} | s \in \Sset_{k}} \alpha_{h}^{s,k} = 1, \mbox{   } \forall i \in \Pset_{s}
\end{align}
Similarly to Eq.~(\ref{opt:intermediate_problem}), we also use the proximal method \cite{bertsekas_parallel_dist_comp_book} to solve the optimization problem in Eq.~(\ref{opt:eq1_trafficSplit}).

\textbf{Scheduling.} Third, we solve the Lagrangian for $\tau_{h}$. This problem is solved for every hyperarc and every clique in the conflict graph in the hypergraph.
\begin{align} \label{opt:eq1_scheduling}
\max_{\boldsymbol \tau} & \sum_{h \in \Aset}  q_{h} R_{h} \tau_{h} \nonumber \\
\mbox{s.t. } & \sum_{h \in \Cset_{q}} \tau_{h} \leq \tau, \mbox{  } \forall \Cset_{q} \subseteq A.
\end{align}

\textbf{Parameter (Queue Size) Update.} We find $q_{h}$, using a gradient descent algorithm: $q_{h}(t+1) = \{ q_{h}(t) + c_t [ \sum_{k \in \Kset_{h}} \sum_{s \in \Sset_{k}} H_{h}^{s,k} \alpha_{h}^{s,k} (m_{h}^{s,k})^{*} x_s  - R_{h} \tau_{h} ] \}^{+}$. Equivalently;
\begin{align} \label{opt:eq1_parameterUpdate}
q_{h}(t+1) = \{ q_{h}(t) + c_t [ \sum_{k \in \Kset_{h}} \max_{s \in \Sset_{k}} \{H_{h}^{s,k} \alpha_{h}^{s,k} x_s\}  - R_{h} \tau_{h} ] \}^{+}
\end{align} where $t$ is the iteration number, $c_t$ is a small constant, and the $~^+$ operator makes the Lagrange multipliers positive.
$q_h$ can be interpreted as the queue size at hyperarc $\forall h \in \Aset$. Indeed, in Eq.~(\ref{opt:eq1_parameterUpdate}), $q_h$ is updated with the difference between the incoming
 $\sum_{k \in \Kset_{h}} \max_{s \in \Sset_{k}} \{H_{h}^{s,k} \alpha_{h}^{s,k} x_s\}$  and outgoing $R_{h} \tau_{h}$ traffic at $h$.
Therefore, we call $q_h$ the hyperarc-queue, or h-queue for brevity.
We confirmed the convergence of $q_h$'s via numerical calculations as seen in Appendix A. 

\section{\label{sec:algs}Network Coding-Aware Implementation}

In the previous section, we saw that the NUM problem decomposed into Eq.~(\ref{opt:eq1_rateControl1}), Eq.~(\ref{opt:eq1_trafficSplit}), Eq.~(\ref{opt:eq1_scheduling}), Eq.~(\ref{opt:eq1_parameterUpdate}), each of which has an intuitive interpretation. In this section, we mimic the properties of the optimal solutions to these problems and propose modifications to the corresponding protocols to make them network coding-aware. It turns out that only changes to queue management at intermediate nodes are crucial, while TCP and scheduling can remain intact. This makes our proposal amenable to practical deployment.

\subsection{Queue Management at Intermediate Nodes (NCAQM)}

\subsubsection{Summary of Proposed Scheme}

We refer to our {\em Network Coding-Aware Queue Management} scheme as NCAQM. NCAQM builds on and extends COPE \cite{cope}. Its goal is to interact with TCP congestion control in such a way that it matches the rates of TCP flows coded together and thus increases network coding opportunities. It achieves this goal through the following minimal changes at intermediate nodes. First, NCAQM stores coded packets in the output queue $\Qset_{i}$, as opposed to COPE that stores uncoded packets. Second, NCAQM maintains state per hyperarc queue $q_h$ and per network code transmitted over each hyperarc $k \in \Kset_{h}$; this is feasible in the setting of wireless mesh with limited number of flows. Third, during congestion, packets are dropped from the flow  that has the largest number of packets, where this number is computed only over h-queues where the flow is dominant. Consider several flows coded together in the same code: the rate of the dominant flow is the rate of the code; and dropping from the dominant flow matches the rates, as desired. We note that intermediate nodes do already network coding operations and can be naturally extended to implement these changes.

\subsubsection{Detailed Description of Proposed Scheme}~~~~

\textbf{Maintaining Queues:}
In \cite{cope}, a wireless node $i$ stores all packets uncoded 
in a single output queue $\Qset_{i}$ and takes decisions at every transmission opportunity about whether to code some of these packets together or not. In contrast, we propose to network code packets, if an opportunity exists, at the time we store them in the queue. Motivated by the fact that Lagrange multiplier (h-queue) $q_h$ in Eq.~(\ref{opt:eq1_parameterUpdate}) can be interpreted as the queue size at hyperarc $h$, we maintain h-queue virtually\footnote{We maintain a virtual, not a physical, h-queue, because the latter would be difficult in practice: (i) the total buffer size is limited and allocating it to h-queues is another control parameter; (ii) h-queues may change over time depending on changes in the topology and traffic scenario; (iii) storing packets in h-queues may reduce network coding opportunities in a packet-based system (although it is optimal in a flow-based system) due to opportunistic network coding.} for each hyperarc at every node, which  keeps track of packets that are network coded and broadcast over $h$. The size of an h-queue is $Q_h$ and how it is determined in practice will be explained later. Each node $i$ maintains a single physical output queue, $\Qset_{i}$, which stores all packets (coded and uncoded depending on the opportunities) passing through it.

\textbf{Network Coding (Alg. 1):}
Motivated by the fact that the incoming traffic in Eq.~(\ref{opt:eq1_parameterUpdate}) is the sum of the network coded flows over $h$, we code packets when they are inserted to output queues. If a network coding opportunity does not exist when the packet arrives at node $i$, we just store it in $\Qset_{i}$ in a FIFO way. Periodically, Alg.~\ref{alg:NC} runs to check all packets in the queue for network coding.

Let $\Qset_{i} = \{p_1, p_2, ..., p_l\}$ where $p_1$ is the first and $p_{l}$ is the last packet in the queue; $l\le L$, where $L$ is the buffer size, {\em i.e.,} the maximum number of packets that can be stored in $\Qset_{i}$. First, $p_1$ is picked for network coding. Since $\Qset_{i}$ stores network coded packets, $p_1$ may be already coded. Independently of whether $p_1$ is network coded or not, it can be further coded with other packets in the queue beginning from $p_2$, if the following two conditions are satisfied; (i) the packets constructing $p_1$ and $p_2$ should be from different flows, and (ii) $p_1 \oplus p_2$ should be decodable at the next hop of all packets that construct the network code. If these conditions are satisfied, we say that the network code is an eligible network code, and $p_1$ is replaced by $p_1 \oplus p_2$. Then $p_1 \oplus p_3$ is checked for network coding, etc. After all packets are checked for network coding, the output queue $\Qset_{i}$ is updated: (i) the final packet $p_1$ is stored in the first slot of the output queue, and (ii) the memory allocated to other packets are freed. Then, the same algorithm is run for packet $p_2$, etc. When a transmission opportunity arises, the first packet from the output queue is checked for network coding again and broadcast over the hyperarc. 

\begin{algorithm}[t!]
 \caption{Network coding in output queue $Q_i$ at node $i$} \label{alg:NC}
\begin{algorithmic}[1]
\begin{footnotesize}
\FOR {$m=1...L$}
\IF{$\exists p_m \in \Qset_{i}$}
\FOR {$n=(m+1)...L$}
\IF{$p_m \oplus p_n$ is eligible}
\STATE $p_m \leftarrow p_m \oplus p_n$
\ENDIF
\ENDFOR
\ENDIF
\STATE Update $\Qset_{i}$
\ENDFOR
\end{footnotesize}
\end{algorithmic}
\end{algorithm}

\begin{algorithm}[t!]
 \caption{Packet dropping at node $i$ during congestion} \label{alg:packet_drop}
\begin{algorithmic}[1]
\begin{footnotesize}
\STATE Initialization: $\Phi _{i}^{s} = 0$, $\forall s \in \Sset$, $\Sset_{i}^{'} = \emptyset$
\IF {$l > L$}
\FOR {$\forall s \in \Sset$}
\STATE Calculate $\Phi _{i}^{s} = \sum_{h(\Jset) | h \in \Aset}Q_{h}\check{w}_{h}^{s}$
\ENDFOR
\STATE $\Sset_{i}^{'} = \arg\max_{s \in \Sset} \{\Phi _{i}^{s}\}$
\STATE Choose a flow $s' \in \Sset_{i}^{'}$ randomly
\IF {$\exists p_n \in \Qset_{i}$, $n=1..l$, from flow $s'$}
\STATE Drop $p_n$
\ELSE
\STATE Drop $p_l$
\ENDIF
\ENDIF
\end{footnotesize}
\end{algorithmic}
\end{algorithm}

Let the number of packets from flow $s$ in node $i$ be $Q_{i}^{s}$. $Q_{i}^{s}$  captures the difference between the incoming and outgoing traffic for flow $s$ at node $i$. Since an h-queue captures the difference between the incoming and outgoing  traffic over a hyperarc, we calculate its size using the following heuristic: $Q_{h} = \sum_{k \in \Kset_{h}} \max_{s \in \Sset_{k}} \{H_{h}^{s,k}\check{\alpha}_{h}^{s,k} Q_{i}^{s}\}$, where $\check{\alpha}_{h}^{s,k}$ is the approximate traffic splitting, explained next.

The traffic splitting parameters $\alpha_{h}^{s,k}$ are found through the optimization problem in Eq.~(\ref{opt:eq1_trafficSplit}).
Through numerical calculations, we made the following observation: each $\alpha_{h}^{s,k}$ converges to the percentage of time that packets from flow $s$ are transmitted with the $k$-th network code over $h$ at node $i$. At each packet transmission, we calculate the probability that a network code $k$ over hyperarc $h$ can be used for flow $s$, over a time window. The average calculated over this window gives a heuristic estimate of the traffic splitting parameter, $\check{\alpha}_{h}^{s,k}$.

\textbf{Packet Dropping (Alg. 2):}
When a node is congested, it decides which packet to drop. In order to eliminate the potential of rate mismatch between flows coded together, we propose that the node compares the number of all (coded and uncoded) packets of each flow, in queues where the flow is dominant ($m_{h}^{s,k}=1$). This is motivated by the optimal rate control in Eq.~(\ref{opt:eq1_rateControl1}). More specifically, for each flow $s$, we calculate $\Phi _{i}^{s} = \sum_{h(\Jset) | h \in \Aset}Q_{h}\check{w}_{h}^{s}$, where $\check{w}_{h}^{s} = \sum_{k \in \Kset_{h} | s \in \Sset_{k}}$  and $H_{h}^{s,k}\check{\alpha}_{h}^{s,k}\check{m}_{h}^{s,k}$. Upon congestion, the $\Phi _{i}^{s}$'s are compared and a packet from the flow with the largest $\Phi _{i}^{s}$ is dropped, preferably the last uncoded  packet. The choice of the last packet is to make it similar to DropTail. The choice of uncoded packet is so as to hurt only one flow, as opposed to several. If there is a tie in the $\Phi$'s between flows, one flow is randomly picked to drop a packet. If all  packets from the selected flow are coded, a new coming packet(s) is dropped instead.

To estimate the dominance indicator $\check{m}_{h}^{s,k}$ needed in Alg.~\ref{alg:packet_drop}, we compute heuristically an estimate $\check{m}_{h}^{s,k}$ as follows. If $H_{h}^{s,k}\check{\alpha}_{h}^{s,k}Q_{i}^{s} < H_{h}^{s',k}\check{\alpha}_{h}^{s',k}Q_{i}^{s'} \text{ s.t. } \exists s' \in \Sset_{k}-\{s\}$, then $\check{m}_{h}^{s,k} = 0$. Otherwise, $\check{m}_{h}^{s,k} = (|\Sset_{k}^{max}|)^{-1}$ where $\Sset_{k}^{max} = \{s| s \in \Sset_{k}$ $\wedge$  $H_{h}^{s,k}\check{\alpha}_{h}^{s,k}Q_{i}^{s}$ $=$ $\max \{H_{h}^{s',k}\check{\alpha}_{h}^{s',k}Q_{i}^{s}$ $|$ $s' \in \Sset_{k} \}\}$.

\subsection{Rate Control at the Sources}
For logarithmic utility, we saw that the optimal rate control in Eq.~(\ref{opt:eq1_rateControl1}) is  $x_s = (\sum_{i \in \Pset_{s}} q_{i}^{s})^{-1}$. $q_{i}^{s}$ corresponds to the length of the network coded queue size of flow $s$ at node $i$. The optimal rate $x_s$ is inversely proportional to the sum of these queue sizes $q_{i}^{s}$ across all nodes $i$ on its path $\Pset_{s}$. This is essentially a generalization of standard optimal rate control \cite{book_srikant}, to account for network coding in the calculation of queue sizes.

When rate control is implemented, it is impractical to feed back to the source the full information $\sum_{i \in \Pset_{s}} q_{i}^{s}$, as required by the optimal control. Instead, when a queue is congested, a packet is dropped or marked  \cite{book_srikant}. The source uses this binary information as a signal to reduce its rate, mimicking the inverse relationship in the optimal control. The exact adaptation of the flow rate depends on the TCP version used. In the simulations, we used TCP-SACK  without any modification. The only change we propose is the packet dropping scheme at the queue (Alg. 2), to take into account not only congestion but also network coding. Essentially, TCP still reacts to drops but these drops are caused when the flow is dominant in at least one network coded queue along the path.

\begin{example}\label{ex2}
Let us re-visit the example in Fig.~\ref{fig:x_topology}. There is only one network coded flow over $h = (I,\{A_2,B_2\})$ and assume that link transmission rates are the same. Then the two flows are always coded together and their traffic splitting parameters approach to $1$. The network coded queue sizes are $\Phi _{I}^{1} = Q_{h}\check{m}_{h}^{1}$ and $\Phi _{I}^{2} = Q_{h}\check{m}_{h}^{2}$, where $Q_h$ is the size of the h-queue for $h = (I,\{A_2,B_2\})$, and $\check{m}_{h}^{1}$ and $\check{m}_{h}^{2}$ are the dominance indicators for the two flows. Since $Q_{h}$ is constant, $\Phi _{I}^{1}, \Phi _{I}^{2}$ depend on $\check{m}_{h}^{1}$ and $\check{m}_{h}^{2}$, {\em i.e.,} on which flow has more packets in the output queue. Upon congestion, a packet from the first is dropped if it has more packets in the queue. Then, $S_1$ will reduce its rate by transmitting less packets, while flow $S_2$ keeps increasing its rate, thus decreasing the probability that there is no packet from the second flow for coding at node $I$. More generally, the interaction of our queue management (NCAQM) mechanism and TCP tends to eliminate the rate mismatch of the flows coded together.
\hfill $\Box$
\end{example}

\subsection{Scheduling}
The scheduling part in Eq.~(\ref{opt:eq1_scheduling}) has two parts: intra- and inter-scheduling that determine which packet to transmit from a node and which node should transmit, respectively. Both  have difficulties in practice. Intra-scheduling causes packet reordering at TCP receivers. Inter-scheduling requires centralized knowledge and it is NP hard and hard to approximate \cite{tutorial_doyle}. Given these difficulties and our original goal to make minimal changes to protocols related to congestion control, we limit our proposed modifications to the queue management. We do not propose new scheduling and we use FIFO scheme for packet transmission and standard 802.11 as wireless MAC.

\begin{figure*}[t!]
\vspace{-0pt}
\begin{center}
\subfigure[Alice-and-Bob Topology]{{\includegraphics[width=5.8cm]{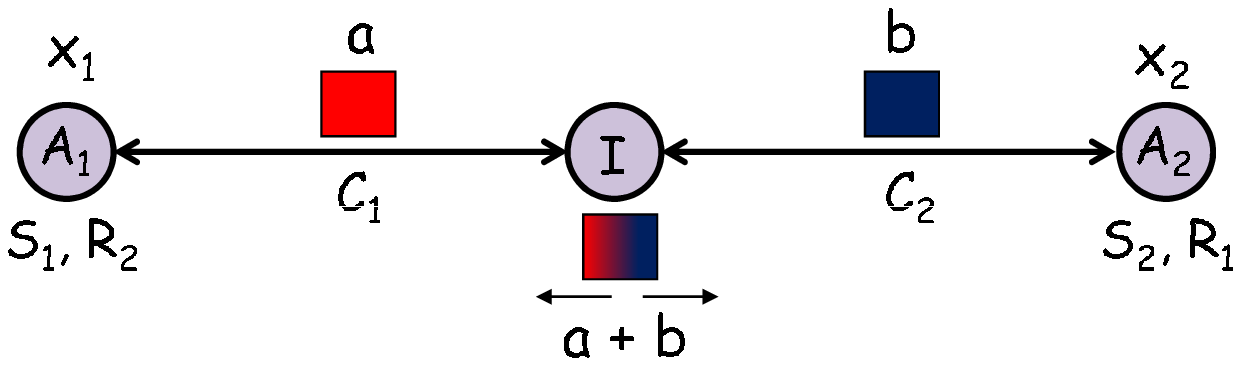}}} \hspace{-0pt}
\subfigure[Cross Topology]{{\includegraphics[width=5.8cm]{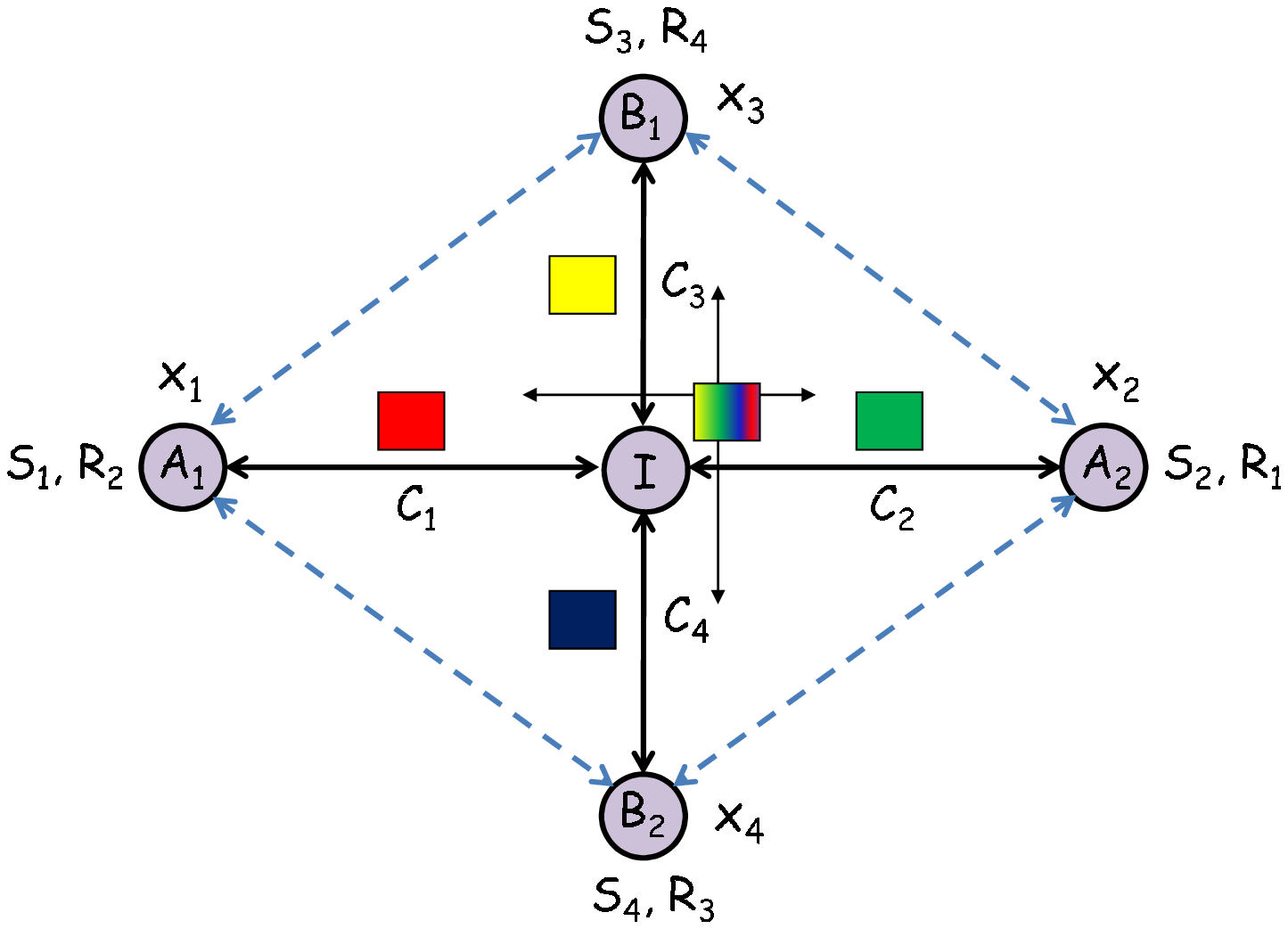}}} \hspace{-0pt}
\subfigure[Wheel Topology]{{\includegraphics[width=5.8cm]{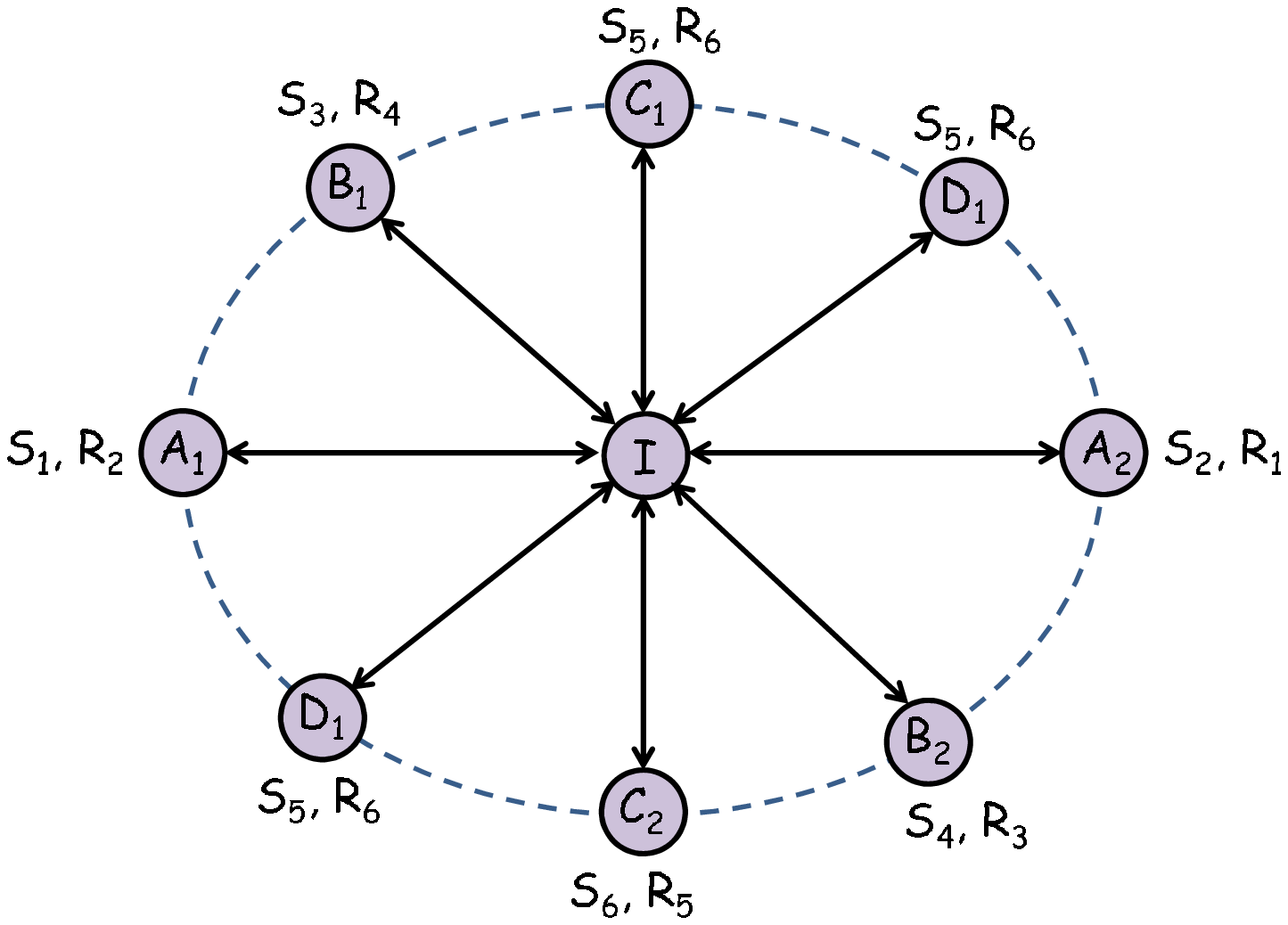}}} \hspace{-0pt}
\end{center}
\begin{center}
\caption{\label{fig:topologies}(a) Alice-and-Bob Topology. Two unicast flows, $S_1-R_1$, and $S_2-R_2$, meeting at intermediate node $I$. (b) Cross topology. Four unicast flows, $S_1-R_1$, $S_2-R_2$, $S_3-R_3$, and $S_4-R_4$, meeting at intermediate node $I$. (c) Wheel topology. Multiple unicast flows $S_1-R_1$, $S_2-R_2$, etc., meeting at intermediate node $I$. $I$ opportunistically combine the packets and broadcast.}
\vspace{-15pt}
\end{center}
\end{figure*}

\begin{figure}
\centering
\includegraphics[width=7cm]{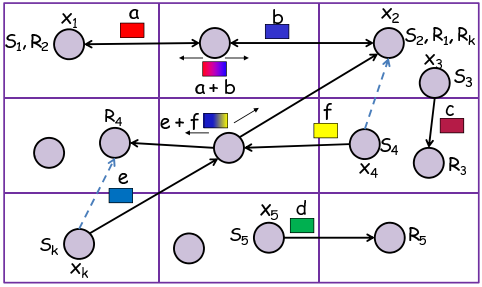}
\caption{Grid topology. Multiple unicast flows $S_1-R_1$, $S_2-R_2$, etc., meeting at different intermediate points.}
\label{fig:grid_topology}
\vspace{-15pt}
\end{figure}

\section{\label{sec:performance}Performance Evaluation}
In this section, we evaluate the throughput of TCP over our proposed scheme (NCAQM) in various topologies and traffic scenarios.
We compare it to TCP over the following baseline schemes: no network coding (noNC), which uses FIFO without network coding; COPE \cite{cope}, which stores native packets in a FIFO and decides which packets to code together at each transmission opportunity; and the optimal control.
\subsection{Simulation Setup}
We used the GloMoSim simulator \cite{glomosim}, which is well suited for wireless. We implemented from scratch the modules for one-hop network coding over wireless mesh networks (COPE) as well as for our proposed scheme (NCAQM). 

\subsubsection{Topologies}
We simulated four illustrative topologies shown in Fig.~\ref{fig:x_topology}, Fig.~\ref{fig:topologies}, and Fig.~\ref{fig:grid_topology}. In X and Alice-and-Bob topologies, shown in Fig.~\ref{fig:x_topology} and Fig.~\ref{fig:topologies}(a), two unicast flows $S_1-R_1$ and $S_2-R_2$ meet at intermediate node $I$. In the cross topology, shown in Fig.~\ref{fig:topologies}(b), four unicast flows $S_1-R_1$, $S_2-R_2$, $S_3-R_3$, and $S_4-R_4$ are transmitted via the relay $I$. In the wheel topology, shown in Fig.~\ref{fig:topologies}(c), multiple unicast flows such as $S_1-R_1$, $S_2-R_2$, $S_3-R_3$, $S_4-R_4$, and etc. are combined at the intermediate node $I$. Note that the wheel topology is the generalized version of the cross topology shown in Fig.~\ref{fig:topologies}(b). In all these topologies node $I$; (i) performs network coding, and (ii) is placed in the center of a circle with $90m$ radius over $200m \times 200m$ terrain and all other nodes are placed around the circle. Finally, we considered the grid topology shown in Fig.~\ref{fig:grid_topology}, in which nodes are distributed over a $300m \times 300m$ terrain, divided into $9$ cells of equal size. $15$ nodes are divided into sets consisting of $1$ or $2$ nodes and each set is assigned to a different cell. Nodes in a set are randomly placed within their cell. If both the transmitter and the receiver are in the same cell or in neighboring cells, there is a direct transmission; otherwise, a node in a neighboring cell acts as a relay. If there are more than one neighboring cells, one is chosen at random. In all topologies, a single channel is used for both uplink and downlink transmissions.

\subsubsection{MAC} In the MAC layer, we simulated IEEE 802.11 with RTS/CTS enabled and with the following modifications for network coding. First, we need a broadcast medium, which is hidden by the 802.11 protocol. We used the pseudo-broadcasting mechanism of \cite{cope}: packets are XOR-ed in a single unicast packet, an XOR header is added for all nodes that should receive that packet, and the MAC address is set to the address of one of the receivers. A receiver knows whether a packet is targeted to it from the MAC address or the XOR header.

\subsubsection{Wireless Channel} We used the two-ray path loss model and Rayleigh fading in Glomosim. We set the average loss rate to 15\%. In our simulations 15\% loss rate is medium loss rate, and residual loss rate after MAC re-transmissions is less than 1\%.\footnote{When channel loss rate increases, there are two problems. First, the residual loss rate after MAC re-transmissions increases. Therefore, TCP is not able to utilize the medium effectively and benefit of network coding reduces. Second, network coding decision at intermediate nodes becomes erroneous, because intermediate nodes do not know which packets are overheard correctly. These issues are out of scope of this paper, and we have analyzed them separately in \cite{hs_intra_inter}.}

\subsubsection{TCP Traffic} We consider FTP/TCP traffic on top of the wireless network. In the Alice-and-Bob, X, cross, and wheel topologies, TCP flows, between the pairs of nodes described above, start at random times within the first $5sec$ and live until the end of the simulation. In the grid topology, TCP flows arrive according to a Poisson distribution with average $6$ flows per $30sec$. The sender and the receiver of a TCP flow are chosen randomly. If the same node is chosen, the random selection is repeated.

\subsection{Simulation Results}
In this section, we present simulation results for the Alice-and-Bob, X, cross, wheel, and grid topologies.  We compare to: (i) TCP over NCAQM (TCP+NCAQM), (ii) TCP over COPE (TCP+COPE), (iii) the optimal solution (optimal rate control in Eq.~(\ref{opt:eq1_rateControl1}) working together with the optimal queue management in Eq.~(\ref{opt:eq1_parameterUpdate})). We report the average throughput of each scheme as \% improvement over the throughput of the baseline TCP+noNC. In addition, we report transport level throughput. All throughput results reported in this section are averaged over $1min$ simulation duration first, then over $10$ simulations with different seeds.

Table~\ref{table:avgThrpt} presents the results for the following parameters: the buffer size at each intermediate node is $10$ packets\footnote{Note that $10$ packet buffer size corresponds to bandwidth-delay product (BDP) in our simulation scenario. We also present simulation results for larger buffer sizes later in this section.}; the packet size is $500B$; the channel capacity is $1Mbps$. In this scenario, TCP+NCAQM has two advantages: (i) it stores network coded packets instead of the uncoded ones, thus uses the buffer more effectively, and (ii) it drops packets so that network coding opportunities increase. Thus, our scheme (TCP+NCAQM) significantly improves throughput as compared to TCP+COPE in all four topologies. It is also seen from the table that there is still a gap between our scheme and the optimal improvement due to the very limited buffer size for multiple flows at the relay. Yet, even in this challenging scenario, TCP+NCAQM significantly improves over TCP+COPE: it doubles the throughput improvement of TCP+COPE.

In Table~\ref{table:avgThrpt}, the improvement of TCP+NCAQM and TCP+COPE in Alice-and-Bob topology is slightly smaller as compared to X topology, although Alice-and-Bob and X topologies have the same optimal improvement (33\%). In Alice-and-Bob topology, source nodes are also receiver nodes, {\em i.e.}, $S_1-R_2$ and $S_2-R_1$ pairs are the same nodes; $A_1$, $A_2$, respectively. Therefore, transport level data and ACK packets share the same buffers at these source/receiver nodes. Due to the limited buffer size, some packets are dropped at the source/receiver nodes, and this reduces TCP throughput. It is also seen that the improvement in cross and grid topologies is larger as compared to Alice-and-Bob and X topologies, for the following reasons: (i) in cross topology, four flows ({\em i.e.}, four packets) are combined at the intermediate node ($I$) instead of two flows, and (ii) in grid topology, we have observed that, during a part of $1min$ simulation duration, four or more flows are combined at intermediate nodes.

\begin{table}
\caption{\label{table:avgThrpt} Average throughput improvement compared to noNC.}
\begin{center}
\begin{tabular}{|c||c|c|c|}
\hline  & {\footnotesize Optimal}  & {\footnotesize TCP+NCAQM} & {\footnotesize TCP+COPE}\\
\hline
{\footnotesize Alice-and-Bob Topology} & {\footnotesize 33\%} & {\footnotesize 18\%} & {\footnotesize 8\%} \\
{\footnotesize X Topology} & {\footnotesize 33\%}  &  {\footnotesize 19\%} & {\footnotesize 9\%} \\
{\footnotesize Cross Topology} & {\footnotesize 60\%} & {\footnotesize 39\%} & {\footnotesize 21\%} \\
{\footnotesize Grid Topology} & {\footnotesize - } & {\footnotesize 35\%} & {\footnotesize 18\%} \\
 \hline
\end{tabular}
\end{center}
\vspace{-10pt}
\end{table}

\begin{figure*}[t!]
\begin{center}
\subfigure[Alice-and-Bob topology]{{\includegraphics[width=6cm]{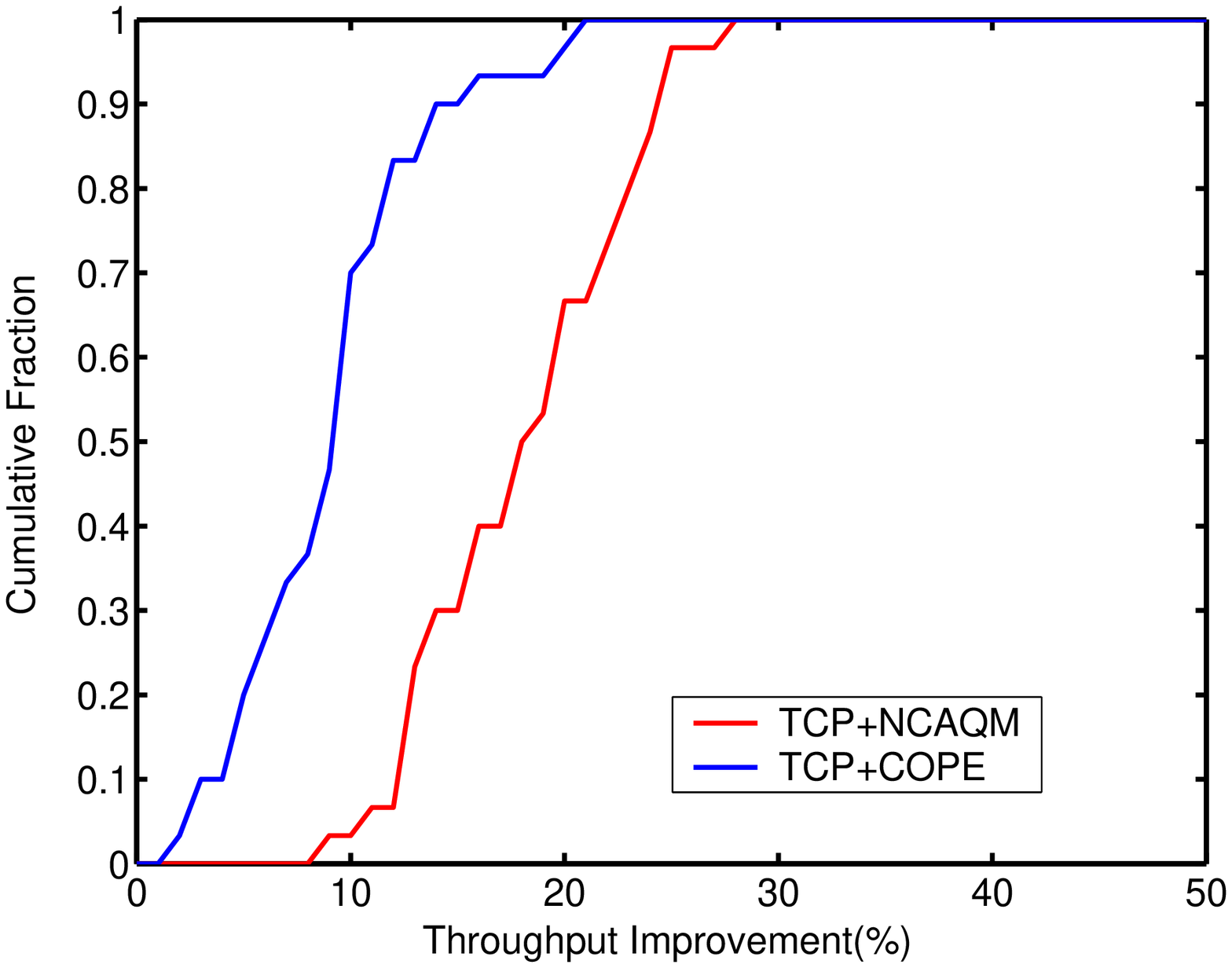}}} \hspace{-0pt}
\subfigure[X topology]{{\includegraphics[width=6cm]{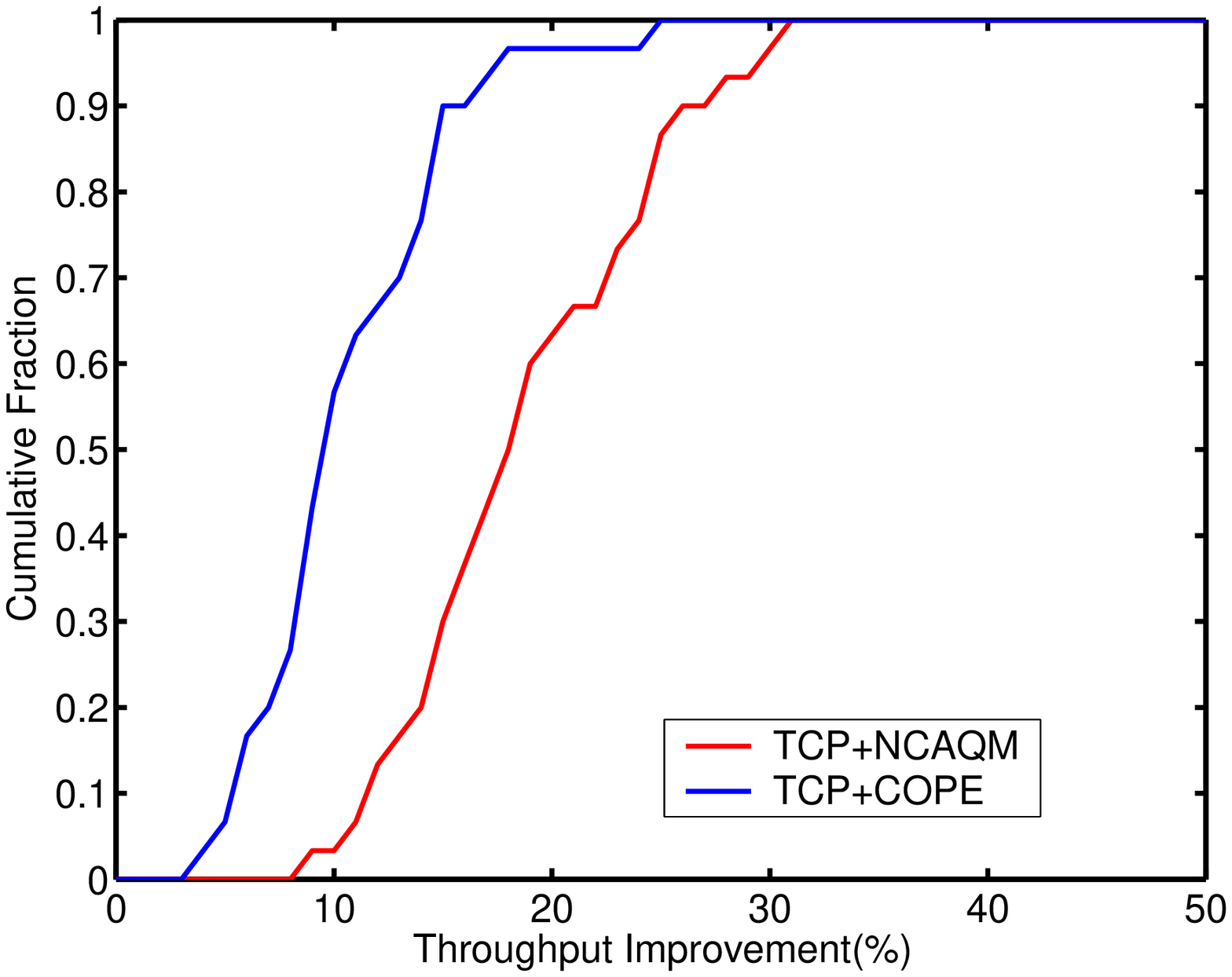}}} \hspace{-0pt}
\subfigure[Cross topology]{{\includegraphics[width=6cm]{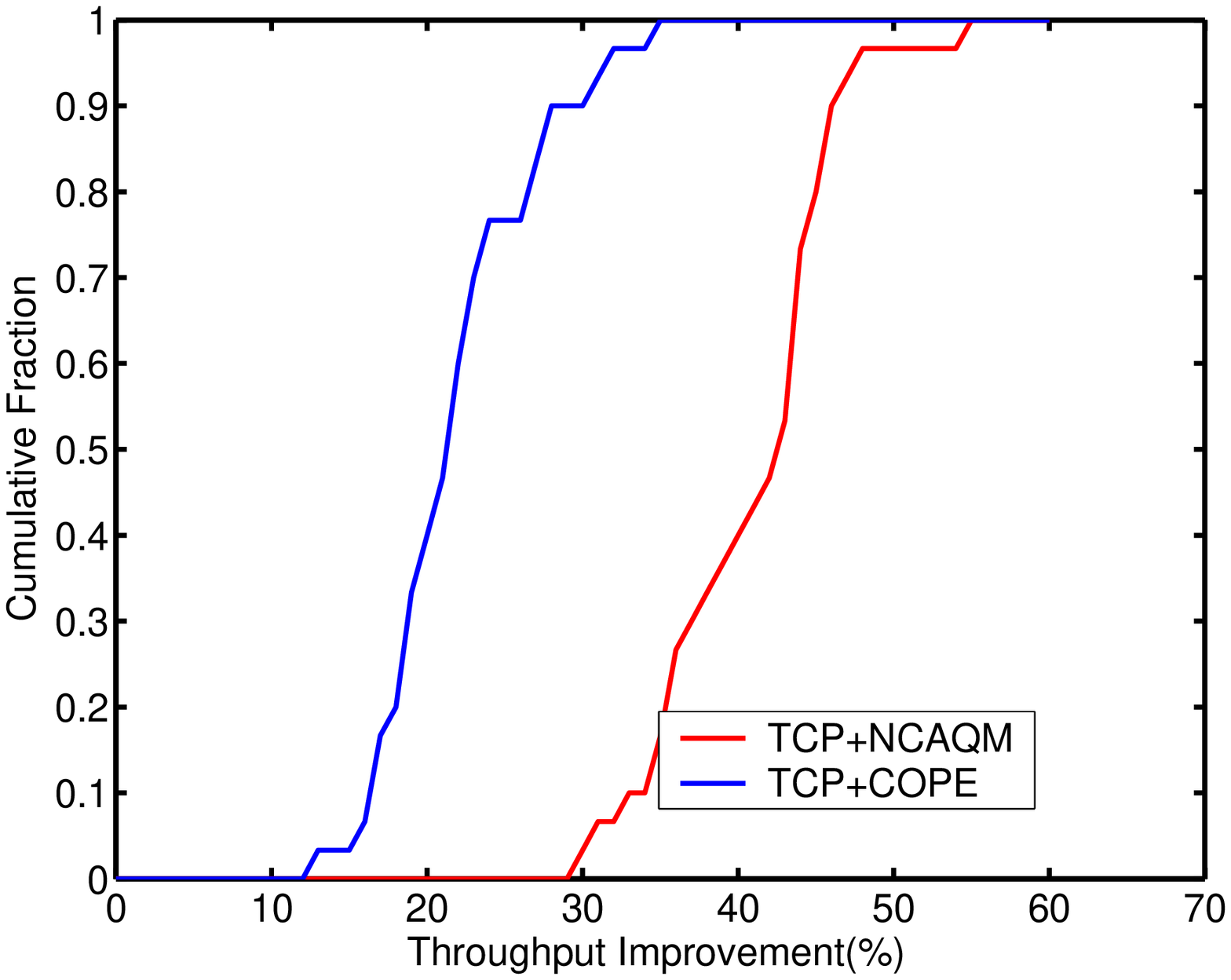}}} \hspace{-0pt}
\subfigure[Grid topology]{{\includegraphics[width=6cm]{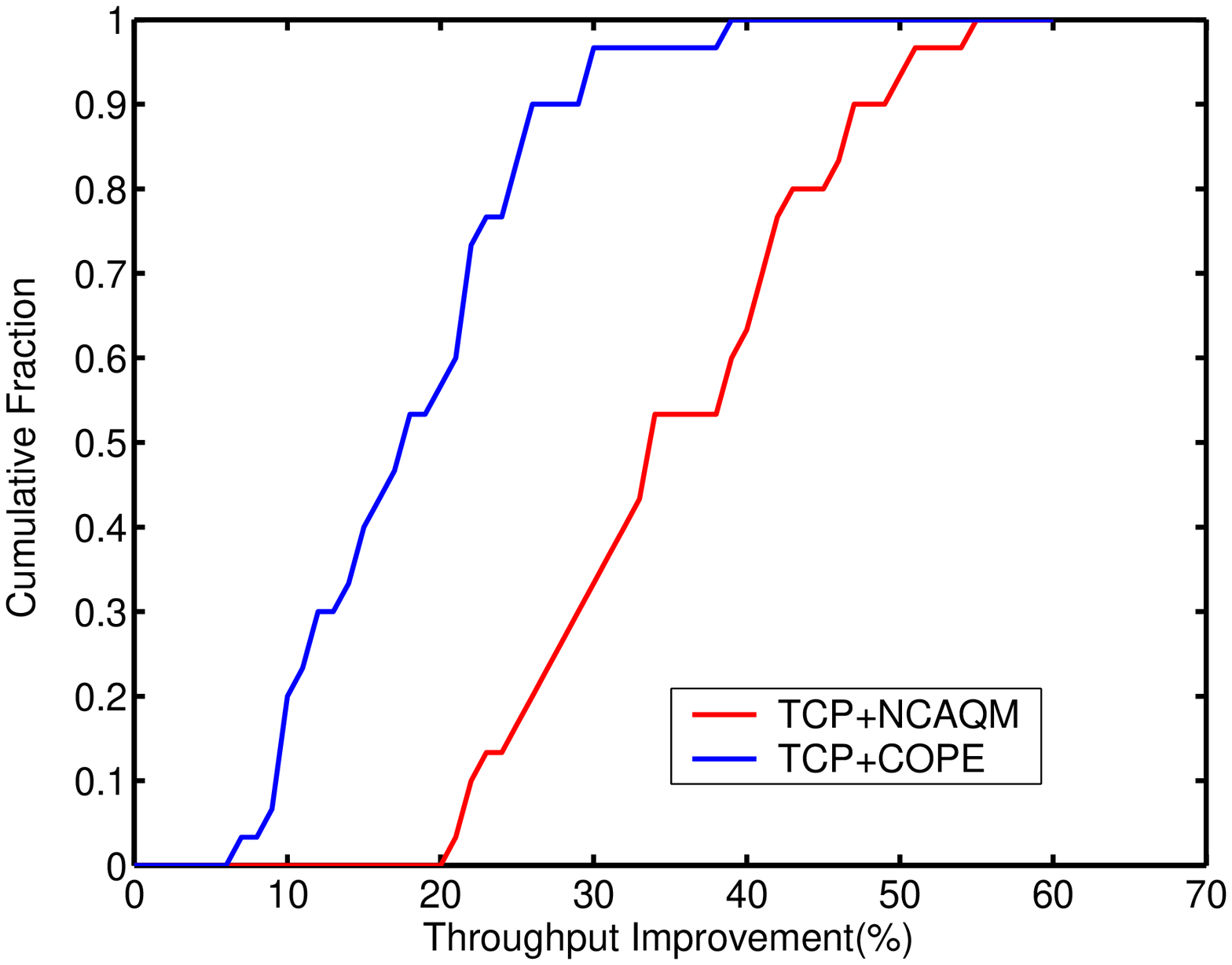}}} \hspace{-0pt}
\end{center}
\begin{center}
\caption{\label{fig:cdfs} Cumulative distribution function (CDF) of throughput improvement for Alice-and-Bob (shown in Fig.~\ref{fig:topologies}(a)), X (shown in Fig.~\ref{fig:x_topology}), cross (shown in Fig.~\ref{fig:topologies}(b)), and grid (shown in Fig.~\ref{fig:grid_topology}) topologies. Buffer size is $10$ packets, packet sizes are $500B$, and the channel capacity is $1Mbps$. The distributions are generated over $30$ seeds.}
\vspace{-20pt}
\end{center}
\end{figure*}

Fig.~\ref{fig:cdfs} presents the cumulative distributed function (CDF) of throughput improvement for the Alice-and-Bob, X, cross, and grid topologies and the same setup. The CDFs are calculated over $30$ seeds. One can see that the CDF of TCP+NCAQM is shifted to significantly higher throughput levels compared to TCP+COPE in all four topologies. For example, TCP+NCAQM improves the throughput more than $20\%$ and $40\%$ in more than $60\%$ of the realizations in Alice-and Bob and cross topologies, respectively. In contrast to Alice-and-Bob and cross topologies, we also observe that the CDF of TCP+NCAQM is shifted to higher throughput levels compared to the CDF of TCP+COPE in the cross and grid topologies. In the cross and grid topologies, it is possible to code more than two flows together, and when the number of flows coded together increases, the way that TCP+NCAQM uses buffers and balances the rates becomes more important. Thus, we see larger improvement in the cross and grid topologies.

\begin{figure*}[t!]
\begin{center}
\subfigure[Alice-and-Bob topology]{{\includegraphics[width=6cm]{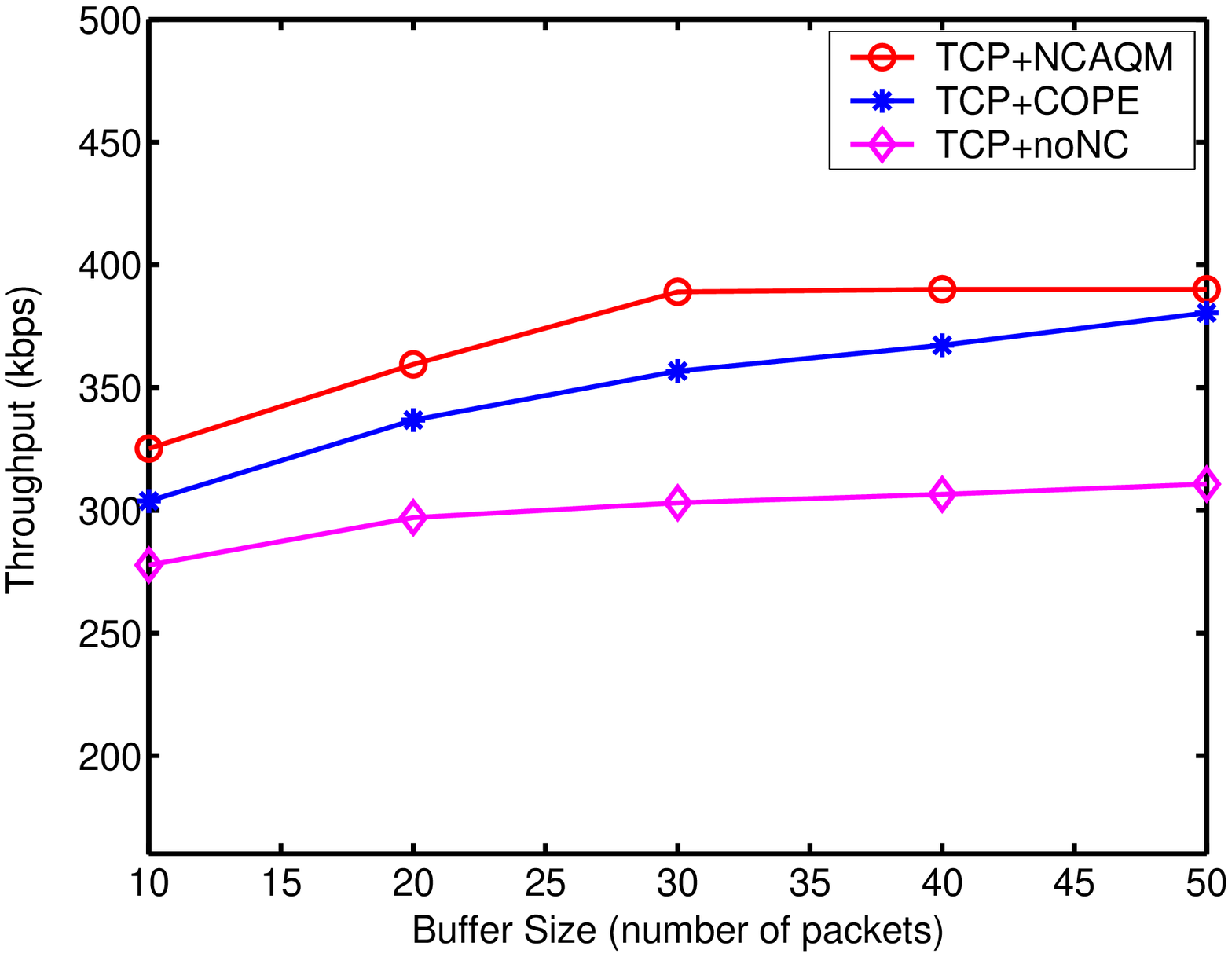}}} \hspace{-0pt}
\subfigure[X topology]{{\includegraphics[width=6cm]{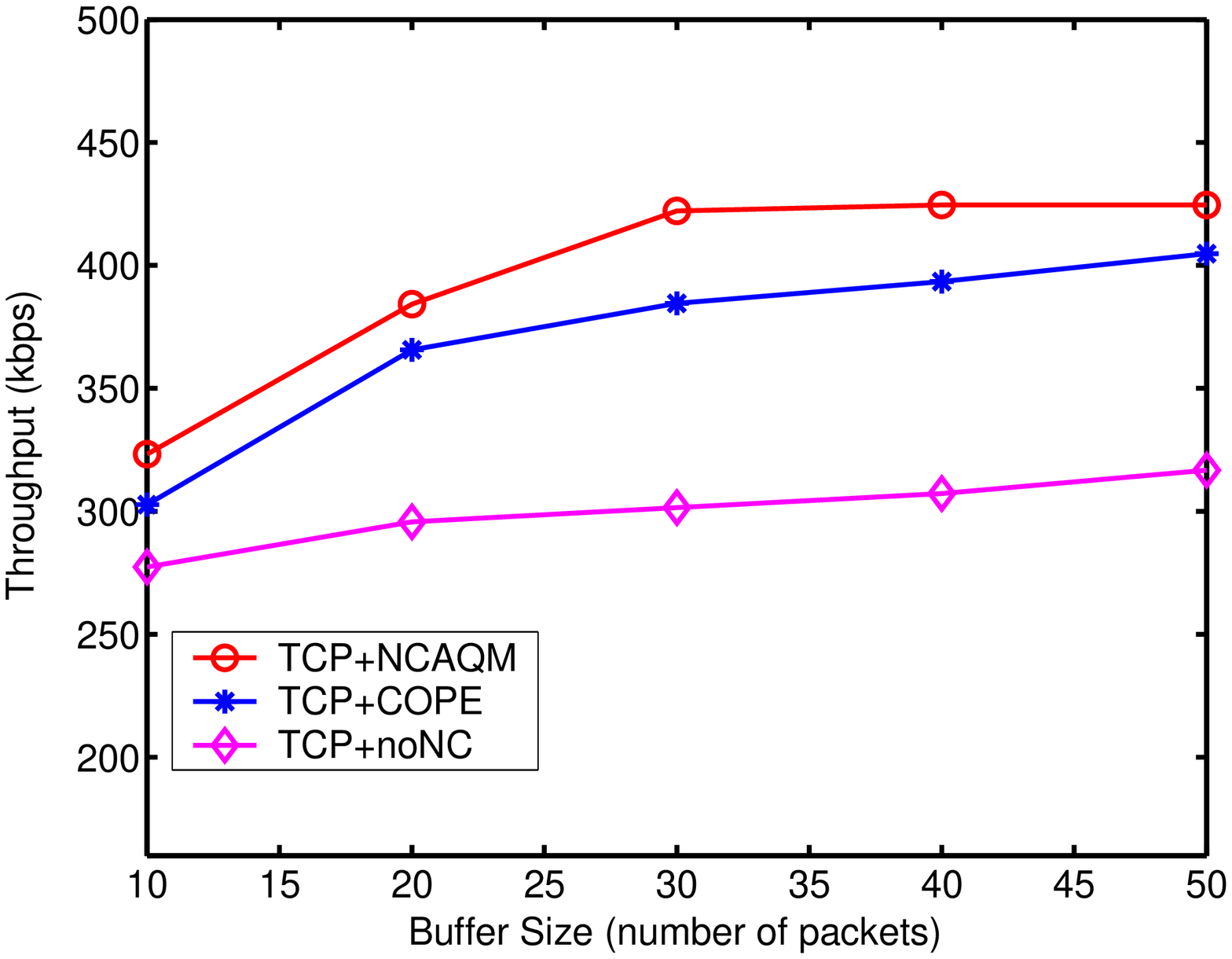}}} \hspace{-0pt}
\subfigure[Cross topology]{{\includegraphics[width=6cm]{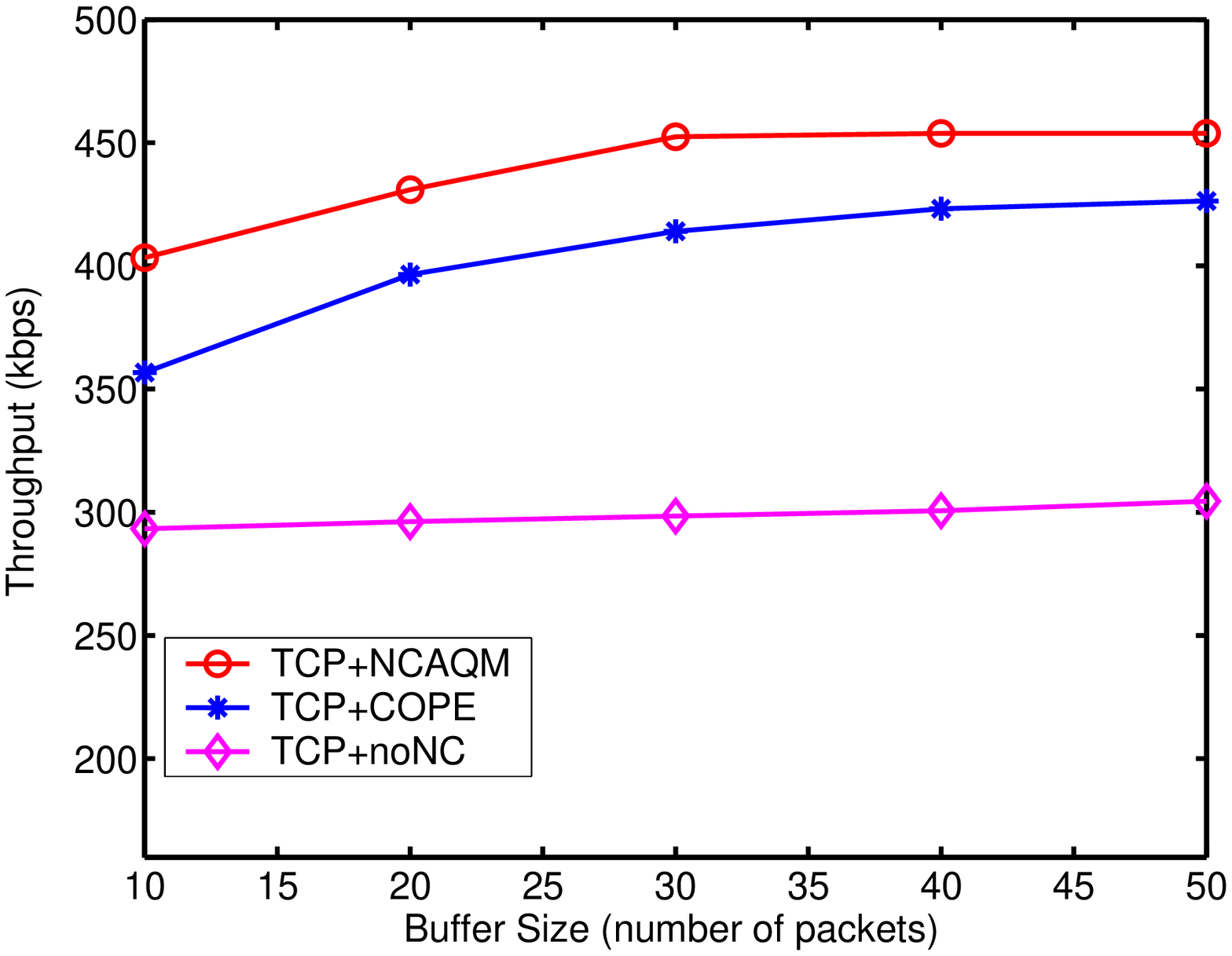}}} \hspace{-0pt}
\subfigure[Grid topology]{{\includegraphics[width=6cm]{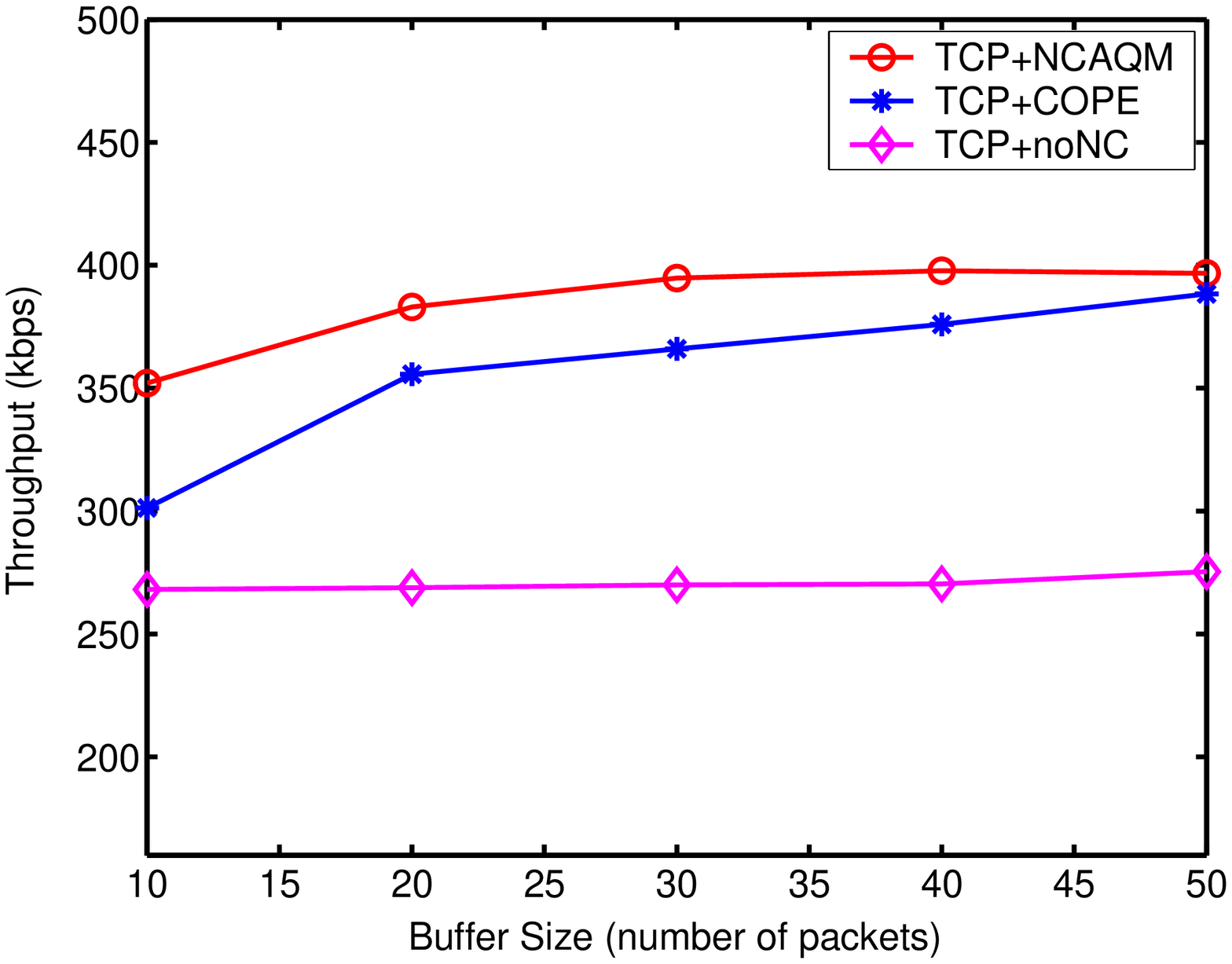}}} \hspace{-0pt}
\end{center}
\begin{center}
\caption{\label{fig:thrpt_vs_buffer_size} Average throughput (averaged in $1min$ simulation first, then over $10$ seeds) versus buffer size for Alice-and-Bob (shown in Fig.~\ref{fig:topologies}(a)), X (shown in Fig.~\ref{fig:x_topology}), cross (shown in Fig.~\ref{fig:topologies}(b)), and grid (shown in Fig.~\ref{fig:grid_topology}) topologies.  Packet size is $500B$, and channel capacity is $1Mbps$.}
\vspace{-20pt}
\end{center}
\end{figure*}

Fig.~\ref{fig:thrpt_vs_buffer_size} shows the average transport-level throughput versus the buffer size, for the Alice-and-Bob, X, cross, and grid topologies. Packet size is $500B$, and channel capacity is $1Mbps$. Our observations from Fig.~\ref{fig:thrpt_vs_buffer_size} are in the following.

The throughput improvement of TCP+noNC for different buffer sizes is negligible in all topologies. The reason is that $10$ packet buffer size is already matched to bandwidth-delay product (BDP) and TCP utilizes wireless medium effectively for almost all buffer sizes when network coding is not used (TCP+noNC). However, for network coding schemes (TCP+NCAQM and TCP+COPE), the throughput increases significantly with increasing buffer size. This shows the importance of active queue management in coded networks.

When buffer sizes are small, the improvement of TCP+NCAQM over TCP+noNC is significantly larger than that of TCP+COPE. This is for the same reason explained earlier: TCP+NCAQM stores network coded, instead of uncoded packets, thus using buffer more effectively, and it drops packets so that network coding opportunities increase. Thus, our scheme (TCP+NCAQM) significantly improves throughput as compared to TCP+COPE in all four topologies.

The throughput of TCP+COPE increases when buffer sizes increase, which is intuitively expected. The problem addressed in this paper was the mismatch between rates of flows coded together, due to the bursty nature of TCP, which reduces coding opportunities. However, when buffer sizes increase, there are more packets available in queues for coding. Thus, TCP+COPE exploits coding opportunities at larger buffers and its throughput increases. However, even at the large buffer sizes, TCP+NCAQM improves throughput more than TCP+COPE. For example, TCP+NCAQM improves throughput 7\% more than TCP+COPE in X topology when buffer size is $50$ packets. Fig.~\ref{fig:thrpt_vs_buffer_size} demonstrates that our scheme is particularly beneficial in harsh buffer size conditions.

The improvement of TCP+NCAQM over TCP+noNC exceeds the optimal throughput at some buffer sizes. {\em E.g.}, the improvement of TCP+NCAQM over TCP+noNC is around 40\% in the X topology when the buffer size is set to $30$ packets (although the optimum improvement is 33\%). The reason is that since TCP+NCAQM uses the buffer more effectively by storing network coded packets instead of uncoded packets, TCP can utilize the medium more effectively, thus the TCP rate increases beyond the network coding benefit.

\begin{figure}
\centering
\includegraphics[width=6cm]{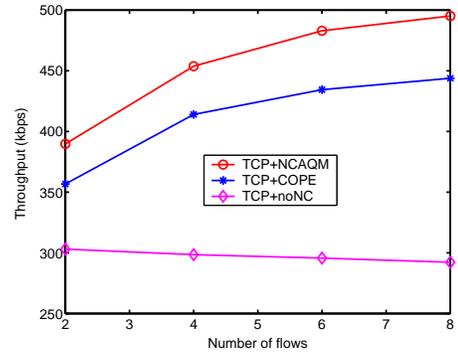}
\caption{Average throughput (averaged in $1min$ simulation first, then over $10$ seeds) versus the number of flows in wheel topology shown in Fig.~\ref{fig:topologies}(c). Buffer size is $30$ packets, packet size is $500B$, and the channel capacity is $1Mbps$.}
\label{fig:cross_numFlows}
\vspace{-15pt}
\end{figure}

Fig.~\ref{fig:cross_numFlows} shows the average transport-level throughput versus the number of flows in the wheel topology shown in Fig.~\ref{fig:topologies}(c). The buffer size is $30$ packets, the packet size is $500B$, and channel capacity is $1Mbps$. One can see from the figure that the throughput of TCP+noNC reduces with increasing number of flows. This is expected, because when the number of flows increases, all flows share the same queue at the intermediate node $I$. As a result, the round trip time of each flow increases, and thus the TCP rate decreases. On the other hand, the throughput of TCP+NCAQM and TCP+COPE increases with the number of flows, because when the number of flows increases, there are more network coding opportunities and more packets can be combined together ({\em i.e.}, it is possible to combine 8 packets when the number of flows is 8). TCP+NCAQM significantly improves over TCP+COPE for all number of flows, especially when the number of flows is large. This is intuitive, because when the number of flows increases, network coding opportunities increases, and TCP+NCAQM exploits these opportunities effectively.

\begin{figure*}[t!]
\begin{center}
\subfigure[Alice-and-Bob topology]{{\includegraphics[width=6cm]{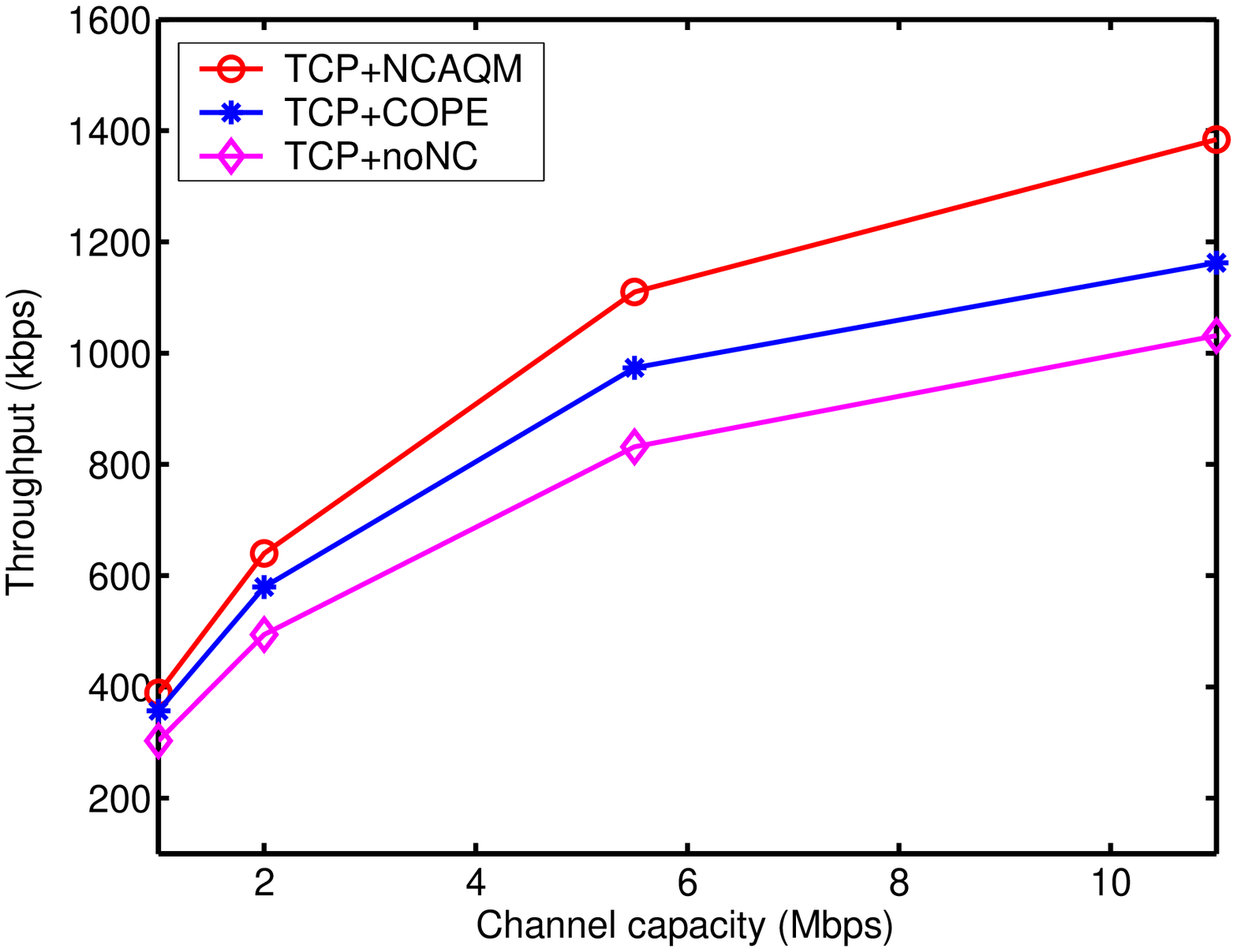}}} \hspace{-0pt}
\subfigure[X topology]{{\includegraphics[width=6cm]{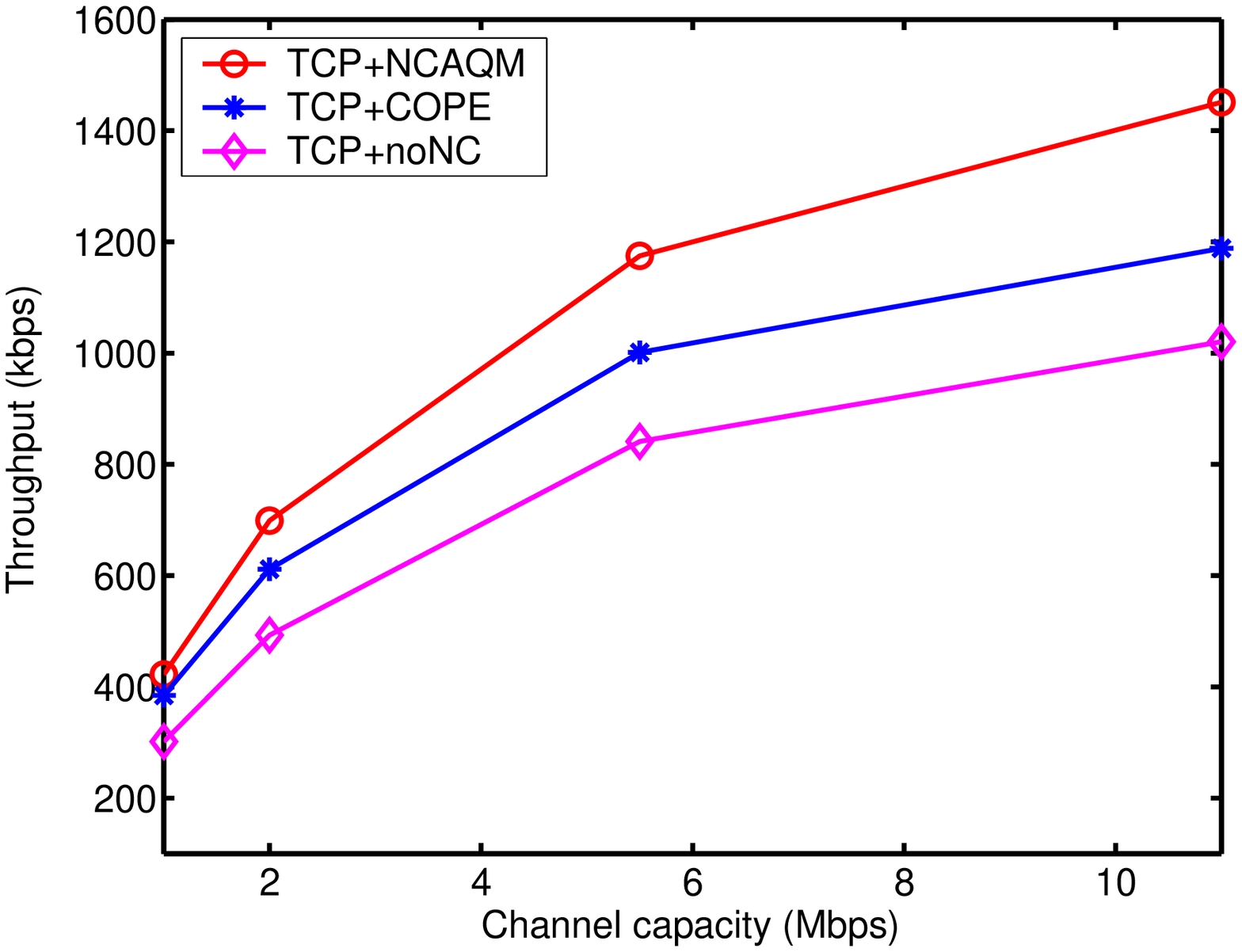}}} \hspace{-0pt}
\subfigure[Cross topology]{{\includegraphics[width=6cm]{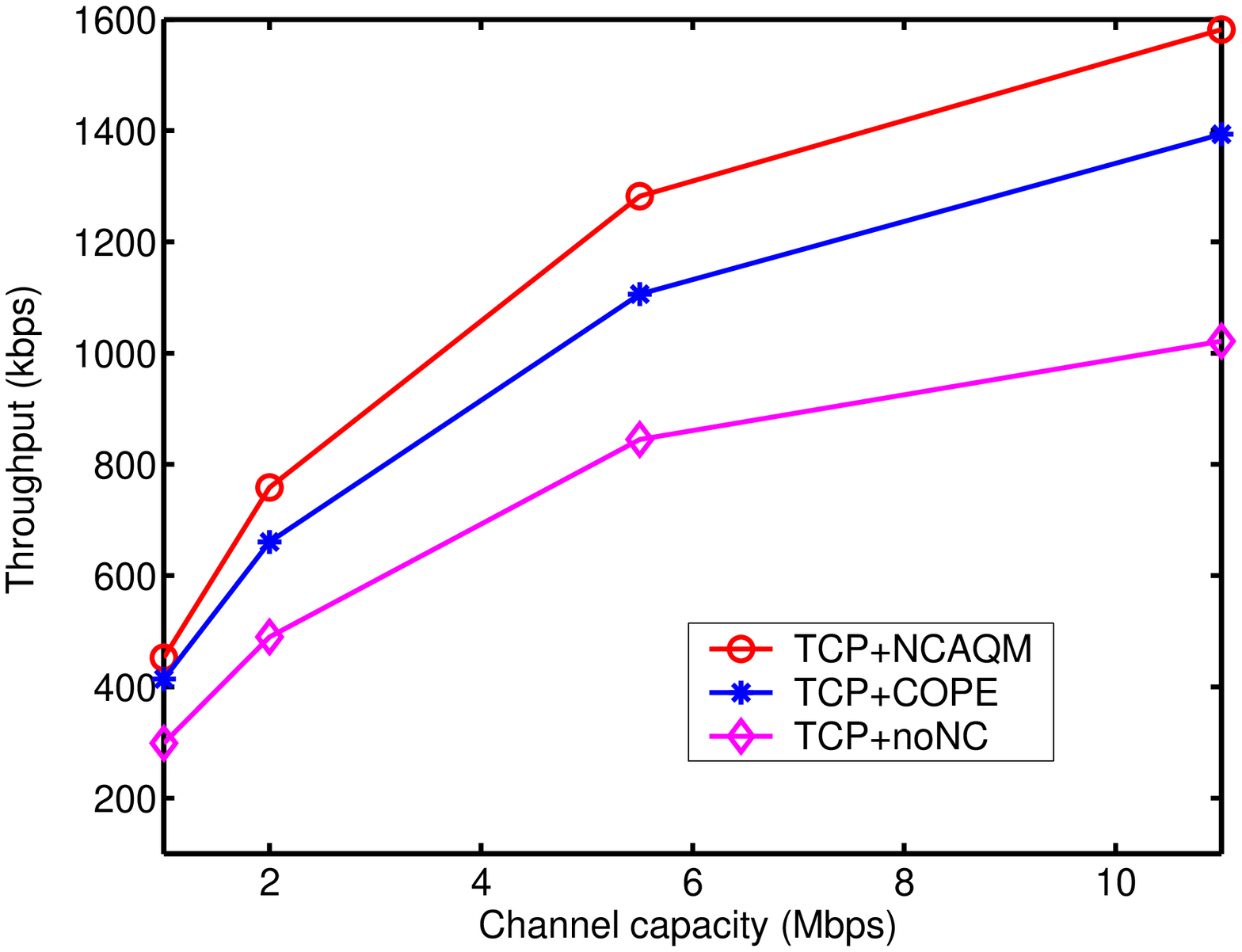}}} \hspace{-0pt}
\subfigure[Grid topology]{{\includegraphics[width=6cm]{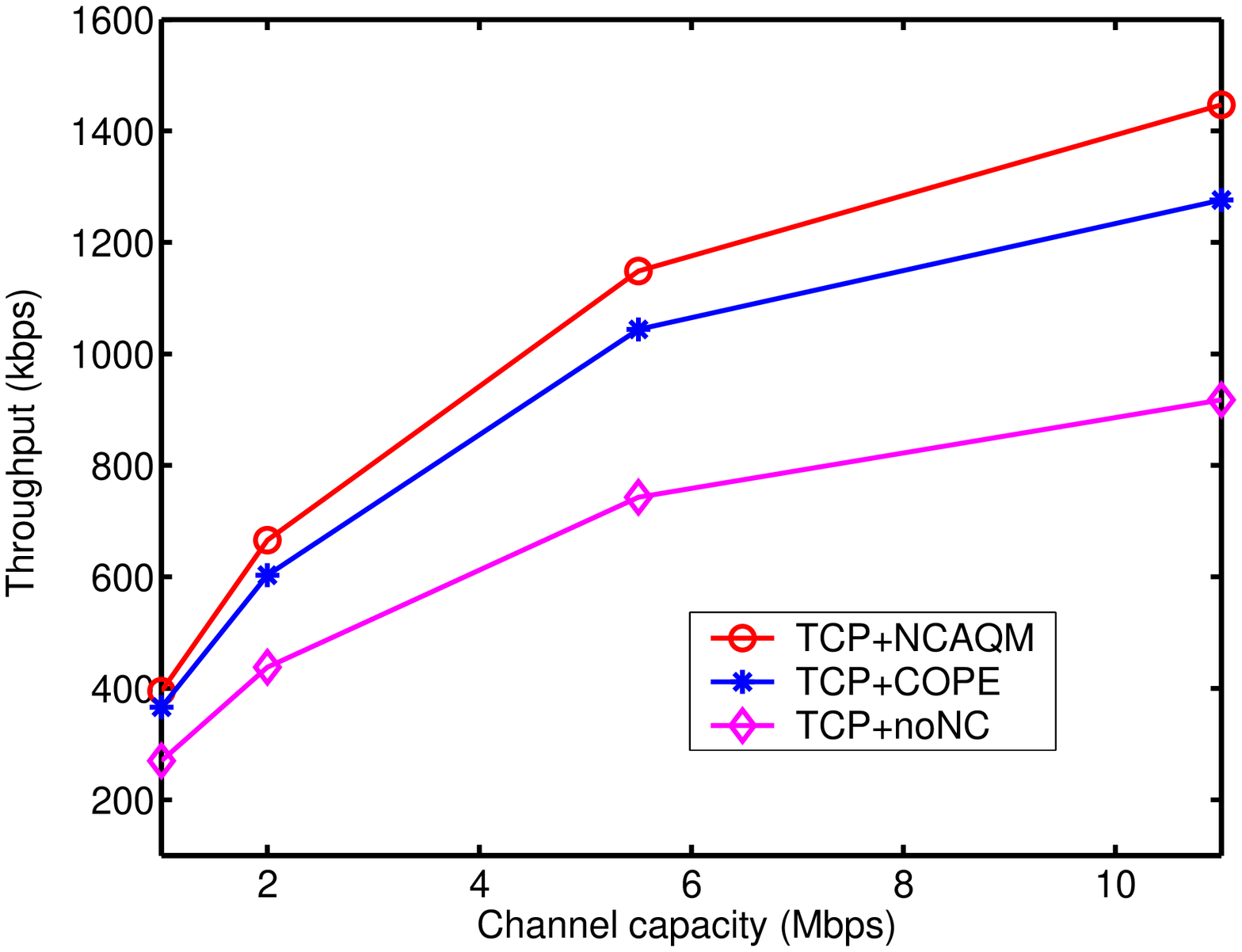}}} \hspace{-0pt}
\end{center}
\begin{center}
\caption{\label{fig:thrpt_vs_bw} Average throughput (averaged in $1min$ simulation first, then over $10$ seeds) versus channel capacity for Alice-and-Bob (shown in Fig.~\ref{fig:topologies}(a)), X (shown in Fig.~\ref{fig:x_topology}), cross (shown in Fig.~\ref{fig:topologies}(b)), and grid (shown in Fig.~\ref{fig:grid_topology}) topologies.  Buffer size is $30$ packets, and packet size is $500B$.}
\vspace{-10pt}
\end{center}
\end{figure*}

Fig.~\ref{fig:thrpt_vs_bw} presents the average transport-level throughput versus channel capacity for the Alice-and-Bob, X, cross, and grid topologies. The buffer size is $30$ packets, and the packet size is $500B$. One can see from the figure that when the channel capacity increases, the gap between TCP+NCAQM and TCP+COPE increases. Therefore, while the improvement of TCP+NCAQM over TCP+noNC increases with increasing channel capacity, it decreases for TCP+COPE. Namely, the improvement of TCP+NCAQM increases from 40\% to 42\%, while the improvement of TCP+COPE decreases from 27\% to 16\% in X the topology. The improvement of TCP+NCAQM is quite significant; more than double the improvement of TCP+COPE at $11Mbps$ channel capacity. The reason is that when the channel capacity increases, more packets share buffer at intermediate node. TCP+NCAQM can improve the throughput by using the shared buffers more effectively, and by dropping packets so as to increase network coding opportunities.

\section{\label{sec:opt_multi_hop}Multi-Hop Network Coding}
\vspace{5pt}
In this section, we extend our framework from one-hop to multi-hop network coding. We note that our framework can accommodate any given multi-hop network coding scheme, but we use BFLY \cite{BFLY} in our simulations, as an example.

\subsection{System Model}
We consider the same system model as in Section~\ref{sec:system}, with the difference of multi-hop, as opposed to one-hop, network coding.
A flow $s$ can be network coded and decoded several times over its path $\Pset_{s}$. The network coded flow may be transmitted over multiple ($M$) hops, which we call $M$-hop network coding. $M$-hop network coding is implemented by COPE \cite{cope} for $M=1$, BFLY \cite{BFLY} for $M=2$, or other network coding schemes for $M>2$. We assume that a flow $s$ cannot be network coded if it (or a part of it) is already coded. This assumption allows  us to divide the path $\Pset_{s}$ to $F_{s}$ intermediate paths which we call {\em network coding paths}. Over its $f$-th network coding path, where $f \in \{1, \ldots, F_{s}\}$, flow $s$ can be network coded with $\Gamma_{f}^{s} \in  \{0, 1, \ldots, |\Sset -\{s\}|\}$ other flows. Without loss of generality, we can assume that a flow may be transmitted over the $f$-th network coding path without network coding; {\em i.e.}, $\Gamma_{f}^{s} = 0$.
 A flow $s$ can be divided into network coded and non-network coded parts over a network coding path $f$, where $\Zset_{s}^{f}$ is the set of partitions of flow $s$ over its $f$-th network coding path. Each partition $z \in \Zset_{s}^{f}$ transmitted over hyperarc $h$ has one-to-one mapping with the $k$-th network code over $h$ such that $k \in \Kset_{h}$, {\em i.e.}, $z=\eta(k)$ over $h$ where $\eta$ is an injective function.

\begin{figure}
\centering \vspace{+30pt}
\includegraphics[width=8cm]{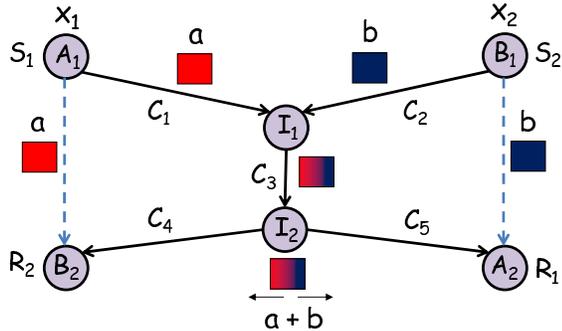}
\vspace{-0pt}
\caption{``Butterfly topology''. Source $S_1$ transmits a flow with rate $x_1$ to receiver $R_1$ and source $S_2$ transmits a flow with rate $x_2$ to receiver $R_2$, over the intermediate nodes $I_1$ and $I_2$. Nodes $A_1$ and $B_1$ transmit their packets $a$ and $b$, in two time slots, and node $I_1$ receives them. Node $B_2$ overhears $a$ and $A_2$ overhears $b$, because $A_1-B_2$ and $B_1-A_2$ are in the same transmission range and they can overhear each other. In the next time slot, $I_1$ transmits the network coded packet $a \oplus b$ to node $I_2$. Finally, $I_2$ broadcast $a \oplus b$ over hyperarc $(I_2,\{A_2,B_2\})$. Since $A_2$ and $B_2$ have overheard $b$ and $a$, they can decode their packets $a$ and $b$, respectively.}
\label{fig:butterfly_topology}
\end{figure}

\begin{example}\label{ex_multi_hop}
The example shown in Fig.~\ref{fig:butterfly_topology} illustrates the problem with 2-hop network coding. The flow from source $S_1$ is transmitted over the link $A_1-I_1$ without network coding and it is network coded over the links $I_1-I_2$ and $I_2-A_2$. Over the network coding path, including the set of nodes $I_1,I_2,A_2$, the flow rate $x_1$ is partitioned into a network coded and a non-network coded part. The network coded part is combined with the corresponding part of the flow from source $S_2$, transmitted over $I_1-I_2$, and broadcast over $(I_2,\{A_2,B_2\})$. The other part is transmitted over $I_1-I_2$ and $I_2-A_2$ without network coding. Similar to the one-hop network coding in Example~\ref{ex1}, if there is a mismatch between the rates $x_1, x_2$ of the two flows, network coding benefit is not fully exploited. The goal is to solve this problem, assuming a given multi-hop network coding scheme.
\hfill $\Box$
\end{example}

\subsection{Problem Formulation}
We consider the following NUM problem;
\begin{align} \label{app2:eq1}
\max_{\boldsymbol x, \boldsymbol \alpha, \boldsymbol \tau} & \sum_{s \in \Sset} U_{s}(x_{s}) \nonumber \\
\mbox{s.t.}  & \sum _{k \in \Kset_{h}} \max_{s \in \Sset_{k}} \{ H_{h}^{s,k} \alpha_{h}^{s,k} x_{s}\} \leq R_{h} \tau_{h}, \mbox{   } \forall \mbox{   } h  \in \Aset \nonumber \\
       & \sum_{z \in \Zset_{s}^{f}} \beta_{f}^{s,z} = 1, \mbox{   } \forall \mbox{   } s \in \Sset, f=1, ..., F_{s}  \nonumber \\
       & \alpha_{h}^{s,k} =
       \begin{cases} \beta_{f}^{s,z}, & \exists \text{ } z = \eta(k), z \in \Zset_{s}^{f}, f=1, ..., F_{s} \\
       0, & \text{otherwise.}
       \end{cases}
       \nonumber \\
       & \sum_{h \in \Cset_q} \tau _{h} \leq \tau,  \mbox{   }  \forall \mbox{   } \Cset_q \subseteq \Aset
\end{align}

The NUM problem in Eq.~(\ref{app2:eq1}) is similar to the one in Eq.~(\ref{opt:eq1}), in terms of the objective functions, capacity and interference constraints. We only need to update the flow conservation constraint (the second constraint) and add the third constraint, as explained below.

We introduce a new traffic splitting parameter $\beta_{f}^{s,z}$ which represents the percentage of the flow rate $x_s$ allocated to the $z$-th partition of flow $s$ over its $f$-th network coding path. The traffic splitting parameters should sum up to $1$ according to the flow conservation constraint over each network coding path (the second constraint). Since there is a one-to-one mapping between the $z$-th partition and the $k$-th network code over $h$, the traffic splitting parameters, $\alpha_{h}^{s,k}$ and $\beta_{f}^{s,z}$ should be equal (the third constraint). This also implies the following equalities; $H_{h}^{s,k} = H_{h}^{s,z}$, $m_{h}^{s,k} = m_{h}^{s,z}$.


\subsection{Solution}
We use  Lagrangian relaxation to solve the optimization problem in Eq.~(\ref{app2:eq1}) by relaxing the capacity constraint with Lagrange multipliers $q_h$. We obtain the same Lagrange function in Eq.~(\ref{opt:eq1_Lagrange1}). The Lagrange function is decomposed into the same subproblems as in Eq.~(\ref{opt:intermediate_problem}), Eq.~(\ref{opt:eq1_rateControl1}), Eq.~(\ref{opt:eq1_scheduling}) and Eq.~(\ref{opt:eq1_parameterUpdate}). The only different subproblem is the traffic splitting problem, which can be expressed as
\begin{align} \label{app2:eq1_trafficSplit}
\min_{\boldsymbol \alpha} & \sum_{h \in \Aset} \sum_{k \in \Kset_{h} | s \in \Sset_{k}} q_{h} H_{h}^{s,k} \alpha_{h}^{s,k} (m_{h}^{s,k})^{*} \nonumber \\
\mbox{s.t. } & \sum_{z \in \Zset_{s}^{f}} \beta_{f}^{s,z} = 1, \mbox{   } \forall \mbox{   } s \in \Sset, f=1, ..., F_{s}  \nonumber \\
             & \alpha_{h}^{s,k} =
             \begin{cases} \beta_{f}^{s,z}, & \exists \text{ } z = \eta(k), z \in \Zset_{s}^{f}, f=1, ..., F_{s} \\
             0, & \text{otherwise.}
             \end{cases}
\end{align}

The objective function in Eq.~(\ref{app2:eq1_trafficSplit}) can be expanded to be $\sum_{f=1}^{F_{s}} \sum_{h \in \Aset^{f}} \sum_{k \in \Kset_{h} | s \in \Sset_{k}} q_{h} H_{h}^{s,k} \alpha_{h}^{s,k} (m_{h}^{s,k})^{*}$, where $\Aset^{f}$ is the set of hyperarcs that originate from the nodes in the $f$-th network coding path of flow $s$. The two objective functions are equivalent considering the fact that the objective function in Eq.~(\ref{app2:eq1_trafficSplit}) is equal to zero for hyperarcs which are not originated from the nodes over the flow's network coding paths, because the indicator functions ($H_{h}^{s,k}$) are zero for those hyperarcs.

Now, let $\Zset_{s}^{f,h}$ represent the set of partitions of the flow $s$ over $h$ in its $f$-th network coding path. Then, $\sum_{k \in \Kset_{h} | s \in \Sset_{k}}$ and $\sum_{z \in \Zset_{s}^{f,h}}$ are equivalent, due to the one-to-one mapping between the $z$-th partition and $k$-the network code over $h$. Usage of $\sum_{z \in \Zset_{s}^{f,h}}$ instead of $\sum_{k \in \Kset_{h} | s \in \Sset_{k}}$ implies the following changes; $\alpha_{h}^{s,k} = \beta_{f}^{s,z}$, $H_{h}^{s,k} = H_{h}^{s,z}$, and $m_{h}^{s,k} = m_{h}^{s,z}$. Then, the problem reduces to
\begin{align} \label{app2:eq2_trafficSplit}
\min_{\boldsymbol \beta} & \sum_{f=1}^{F_s} \sum_{h \in \Aset^{f}} \sum_{z \in \Zset_{s}^{f,h}} q_{h} H_{h}^{s,z} \beta_{f}^{s,z} (m_{h}^{s,z})^{*} \nonumber \\
\mbox{s.t. } & \sum_{z \in \Zset_{s}^{f}} \beta_{f}^{s,z} = 1, \mbox{   } \forall \mbox{   } s \in \Sset, f=1, ..., F_{s}
\end{align}


The objective function in Eq.~(\ref{app2:eq2_trafficSplit}) can be expressed as $\sum_{f=1}^{F_s} \sum_{z \in \Zset_{s}^{f}} \sum_{h \in \Aset^{f,z}} q_{h} H_{h}^{s,z} \beta_{f}^{s,z} (m_{h}^{s,z})^{*}$ where $\Aset^{f,z}$ which is the subset of $\Aset^{f}$ contains the hyperarcs over which the $z$-th partition of the $f$-th network coding path of flow $s$ is transmitted. The two objective functions are equivalent, because the indicator functions ($H_{h}^{s,k}$) are zero for $h \not \in \Aset^{f,z}$. Finally, the traffic splitting problem for $s \in \Sset, f=1, ..., F_s$ is expressed as
\begin{align} \label{app2:eq3_trafficSplit}
\min_{\boldsymbol \beta} & \sum_{z \in \Zset_{s}^{f}} \beta_{f}^{s,z} \left( \sum_{h \in \Aset^{f,z}} q_{h} H_{h}^{s,z}  (m_{h}^{s,z})^{*} \right) \nonumber \\
\mbox{s.t. } & \sum_{z \in \Zset_{s}^{f}} \beta_{f}^{s,z} = 1, \mbox{   } \forall \mbox{   } s \in \Sset, f=1, ..., F_{s}
\end{align} Similar to what we have done to solve Eq.~(\ref{opt:eq1_trafficSplit}), we use the proximal method  \cite{bertsekas_parallel_dist_comp_book} to solve this problem.

\subsection{Simulation Results}
In this section, we evaluate the throughput of TCP over NCAQM compared to TCP over the following baseline schemes: no network coding (noNC), which uses FIFO without network coding; BFLY \cite{BFLY}, which utilizes knowledge of the local topologies by exchanging periodic messages that includes neighbors of nodes and source route information in the packet headers to exploit butterfly structures in wireless mesh networks. Similarly to COPE, BFLY stores native packets in a FIFO and decides which packets to code together at each transmission opportunity. We used the GloMoSim simulator \cite{glomosim} to implement the modules for two-hop network coding over wireless mesh networks (BFLY) as well as for our proposed scheme (NCAQM).

We simulate the butterfly topology shown in Fig.~\ref{fig:butterfly_topology} in which two unicast flows $S_1,R_1$ and $S_2,R_2$ meet at intermediate node $I_1$. In this topology, nodes are placed over $300m \times 300m$ in butterfly like structure and a single channel is used for both uplink and downlink transmissions. We consider the same MAC update and wireless channel model as in Section~\ref{sec:performance}. We consider FTP/TCP traffic over the wireless network. TCP flows, between the pairs of nodes described above, start at random times within the first $5sec$ and live until the end of the simulation.

\begin{figure*}[t!]
\begin{center}
\subfigure[Buffer Size]{{\includegraphics[width=6cm]{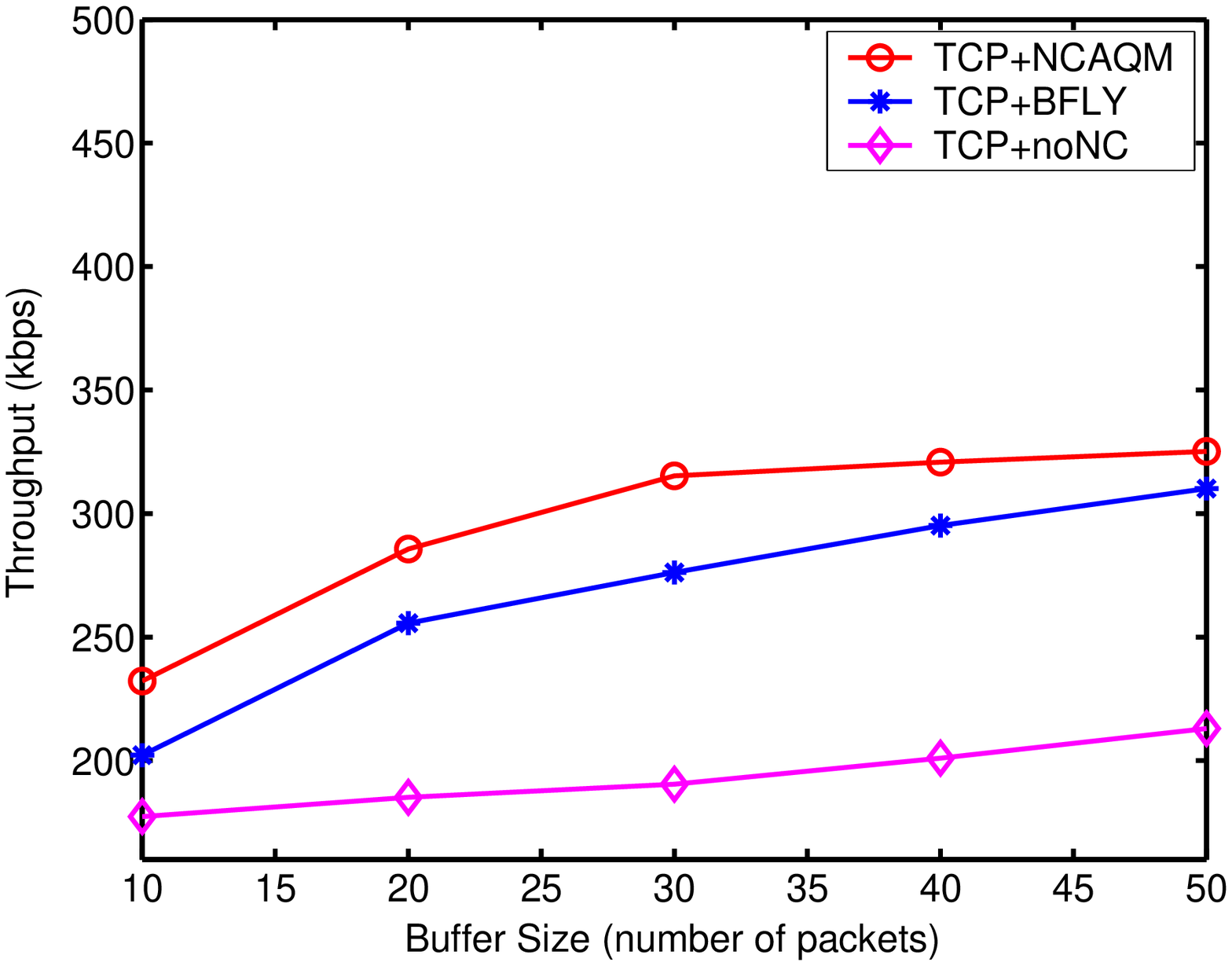}}} \hspace{-0pt}
\subfigure[Channel capacity]{{\includegraphics[width=6cm]{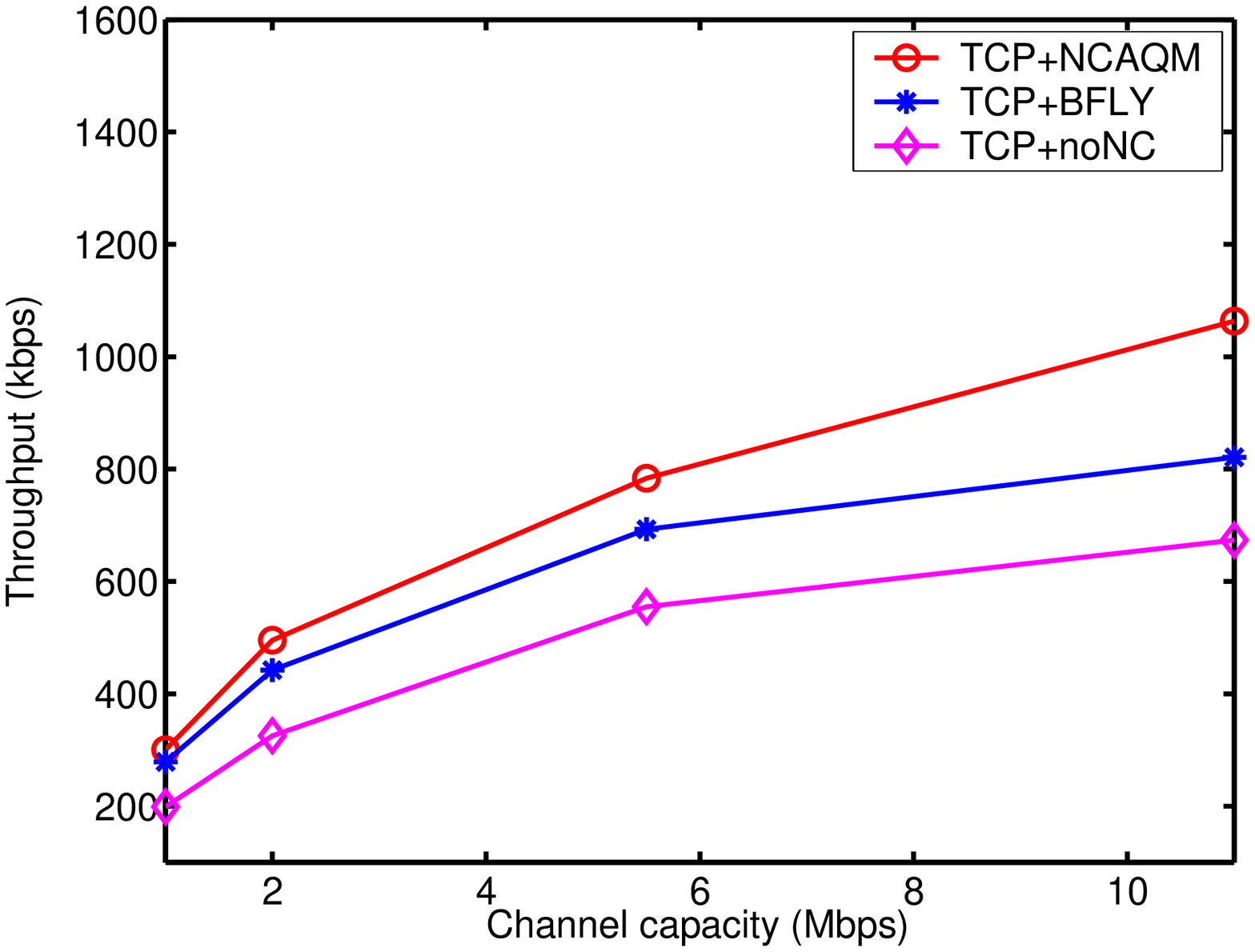}}} \hspace{-0pt}
\end{center}
\begin{center}
\caption{\label{fig:thrpt_vs_buffer_size_bw_butterfly} Average throughput (averaged in $1min$ simulation first, then over $10$ seeds) in butterfly topology shown in Fig.~\ref{fig:butterfly_topology}. (a) Buffer size: packet size is $500B$, and the channel capacity is $1Mbps$. (b) Channel capacity: buffer size is $30$ packets, and packet size $500B$.}
\end{center}
\end{figure*}

Fig.~\ref{fig:thrpt_vs_buffer_size_bw_butterfly}(a) presents the average transport-level throughput vs. buffer size. Similarly to the simulation results in Section~\ref{sec:performance}, TCP+NCAQM improves throughput much more than TCP+BFLY. Specifically, when buffer size is $10$ packets, the improvement of TCP+BFLY over TCP+noNC is 13\%, the improvement of TCP+NCAQM over TCP+noNC is 30\%, while the optimum improvement is 50\%. When buffer size increases, we see that TCP+NCAQM approaches and exceeds the optimum; {\em e.g.}, the improvement of TCP+NCAQM is 65\% when buffer size is $30$ packets, while it is 45\% for TCP+BFLY. This shows that the advantages of TCP+NCAQM also apply to two-hop network coded wireless mesh networks.

Fig.~\ref{fig:thrpt_vs_buffer_size_bw_butterfly}(a) presents the average transport-level throughput vs. channel capacity. We can see that the improvement of TCP+NCAQM is larger than TCP+BFLY for all channel capacities and it is especially significant for large channel capacities, since TCP+NCAQM uses buffer more effectively and  drops packets so that network coding opportunities increase.

\section{\label{sec:conclusion}Conclusion}
In this paper, we showed how to improve the performance of TCP over wireless networks with inter-session network coding. The key intuition was to eliminate the rate mismatch between flows that are coded together through a synergy of rate control and queue management. First, we formulated congestion control as a NUM problem and derived a distributed solution. Motivated by the structure of the solution, we proposed minimal modifications to queue management to make it network coding-aware, while TCP and MAC protocols remained intact. Simulation results show that the proposed NCAQM scheme doubles TCP performance compared to baseline schemes and achieves near-optimal performance. We plan to make the simulator modules publicly available to the research community. We have also extended the NUM formulation and solution to multi-hop network coding and we have confirmed convergence through numerical calculations. The main ideas of this paper can potentially be extended from wireless mesh networks to wired networks with constructive inter-session network coding.

\bibliographystyle{IEEEtran}


\begin{figure*}[t!]
\begin{center}
\subfigure[Rate]{{\includegraphics[width=5cm]{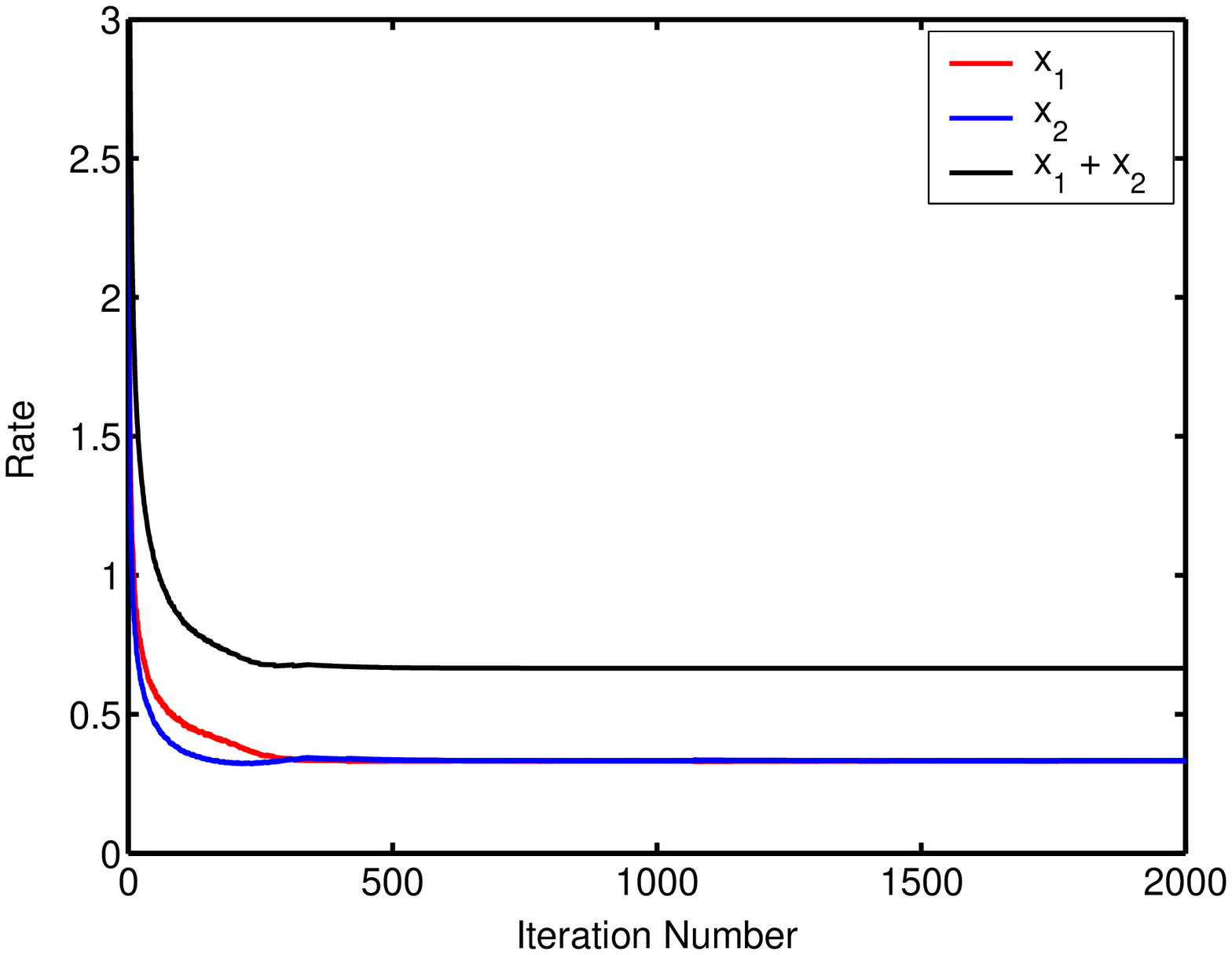}}}
\subfigure[Lagrange multipliers]{{\includegraphics[width=5cm]{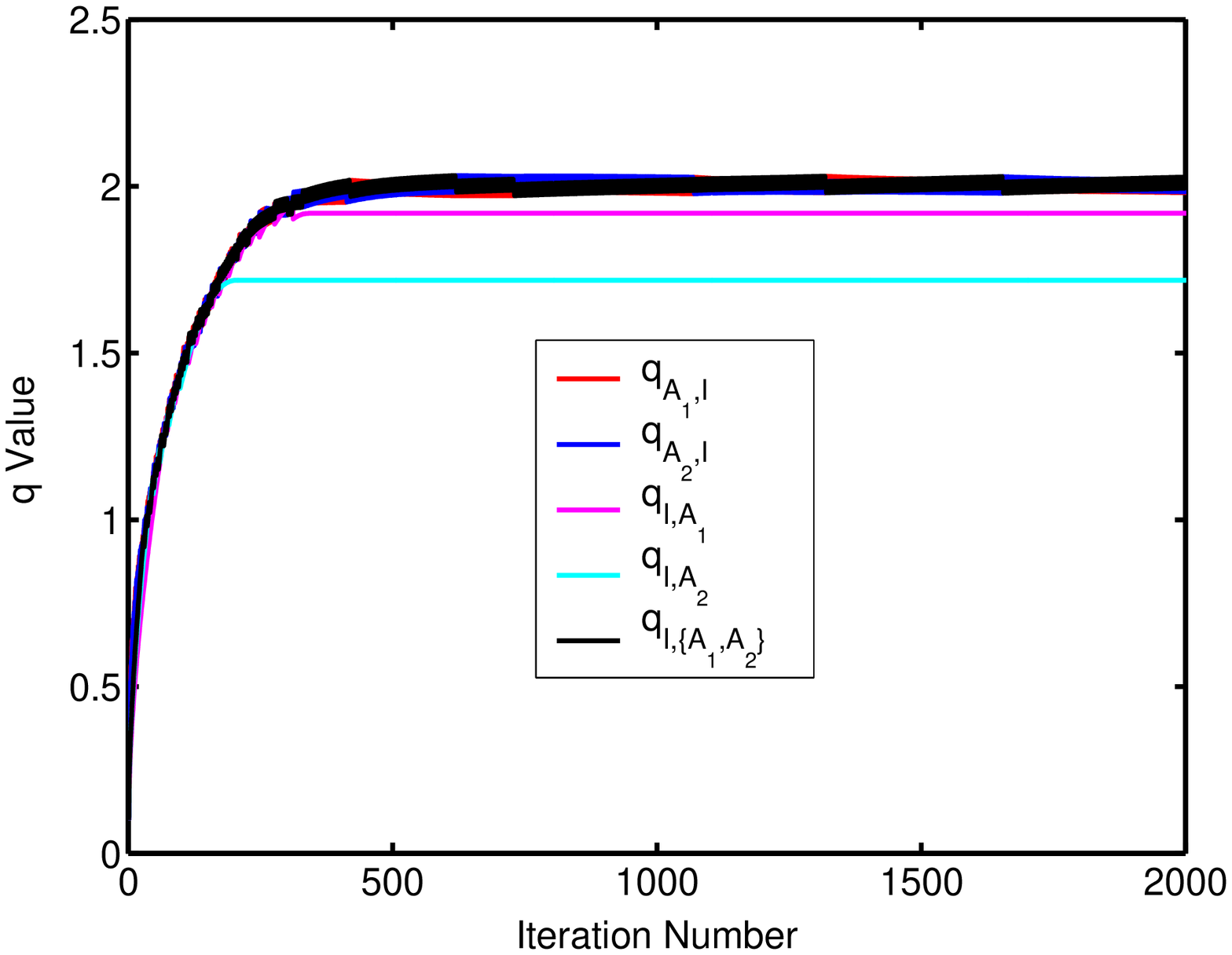}}}
\end{center}
\begin{center}
\caption{Convergence results for the Alice-and-Bob topology presented in Fig.~\ref{fig:topologies}(a). The total achieved rate approaches the optimum throughput 0.66. The optimum throughput is 0.50 when there is no network coding. $C_1=C_2=1$.} \label{fig:num_calc_A_B_case1}
\vspace{5pt}
\end{center}
\end{figure*}

\begin{figure*}[t!]
\begin{center}
\subfigure[Rate]{{\includegraphics[width=5cm]{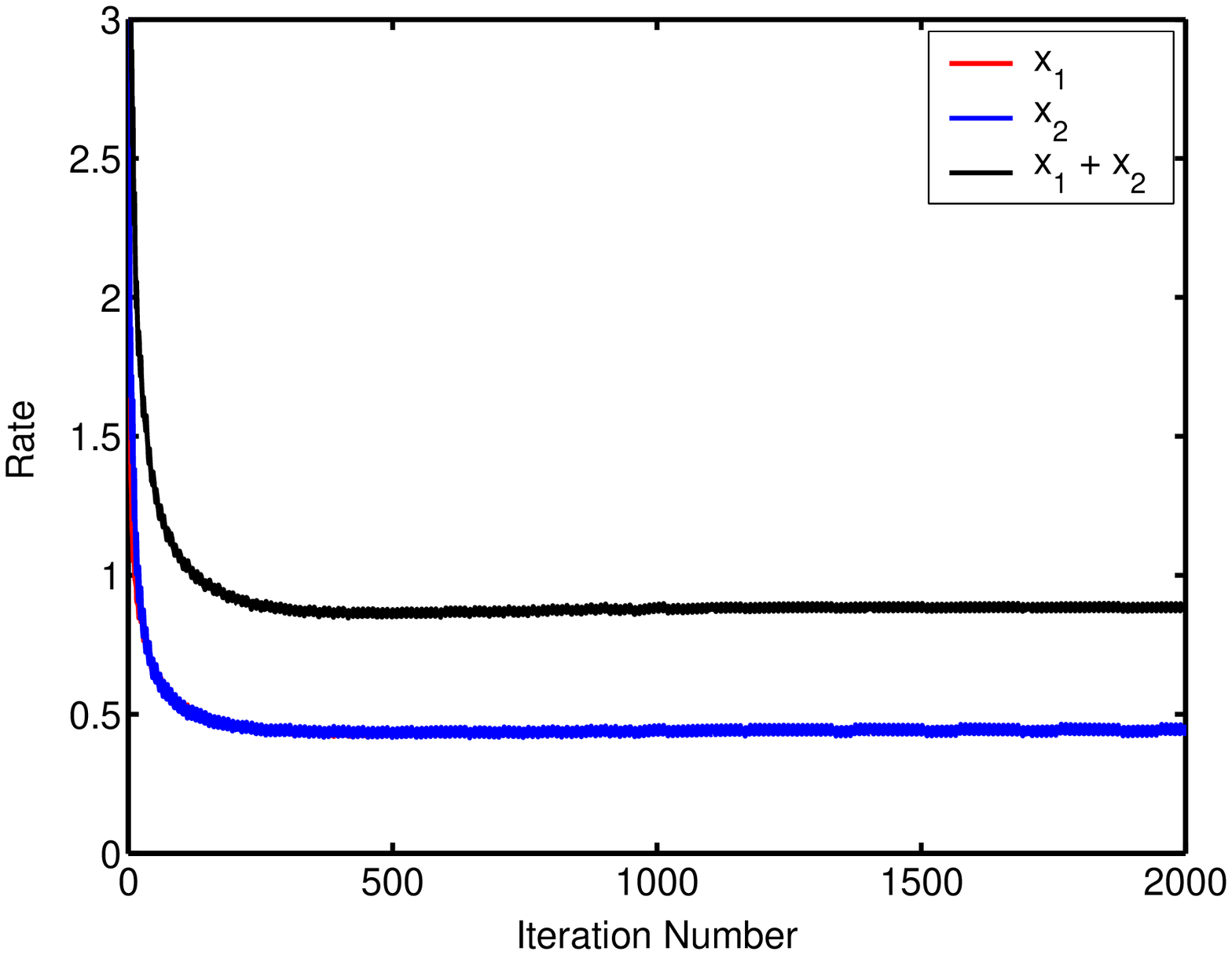}}}
\subfigure[Lagrange multipliers]{{\includegraphics[width=5cm]{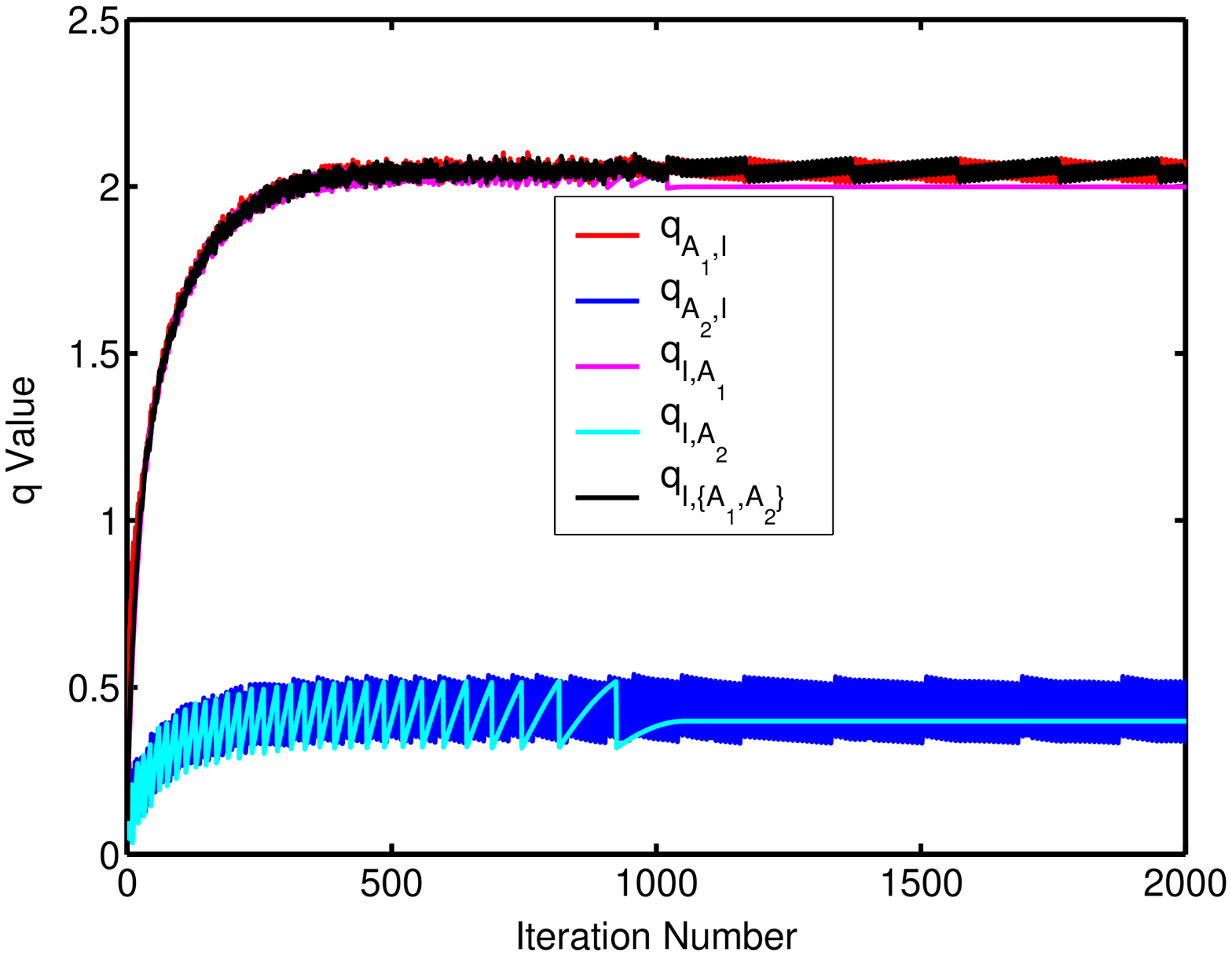}}}
\end{center}
\begin{center}
\caption{Convergence results for the Alice-and-Bob topology presented in Fig.~\ref{fig:topologies}(a). The total achieved rate approaches the optimum throughput 0.88. The optimum throughput is 0.80 when there is no network coding. $C_1=1$, $C_2=4$.} \label{fig:num_calc_A_B_case2}
\vspace{5pt}
\end{center}
\end{figure*}

\begin{figure*}[t!]
\begin{center}
\subfigure[Rate]{{\includegraphics[width=5cm]{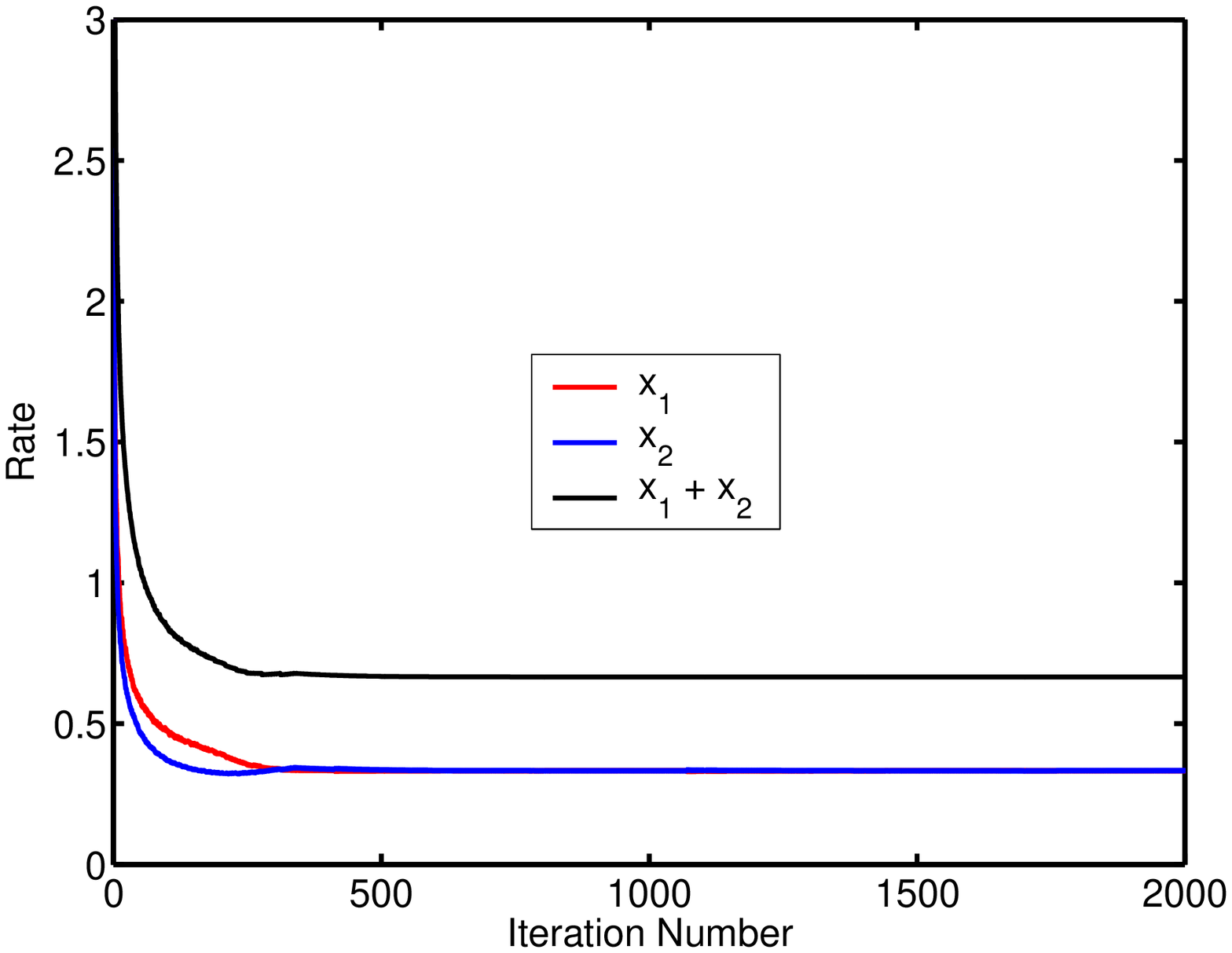}}}
\subfigure[Lagrange multipliers]{{\includegraphics[width=5cm]{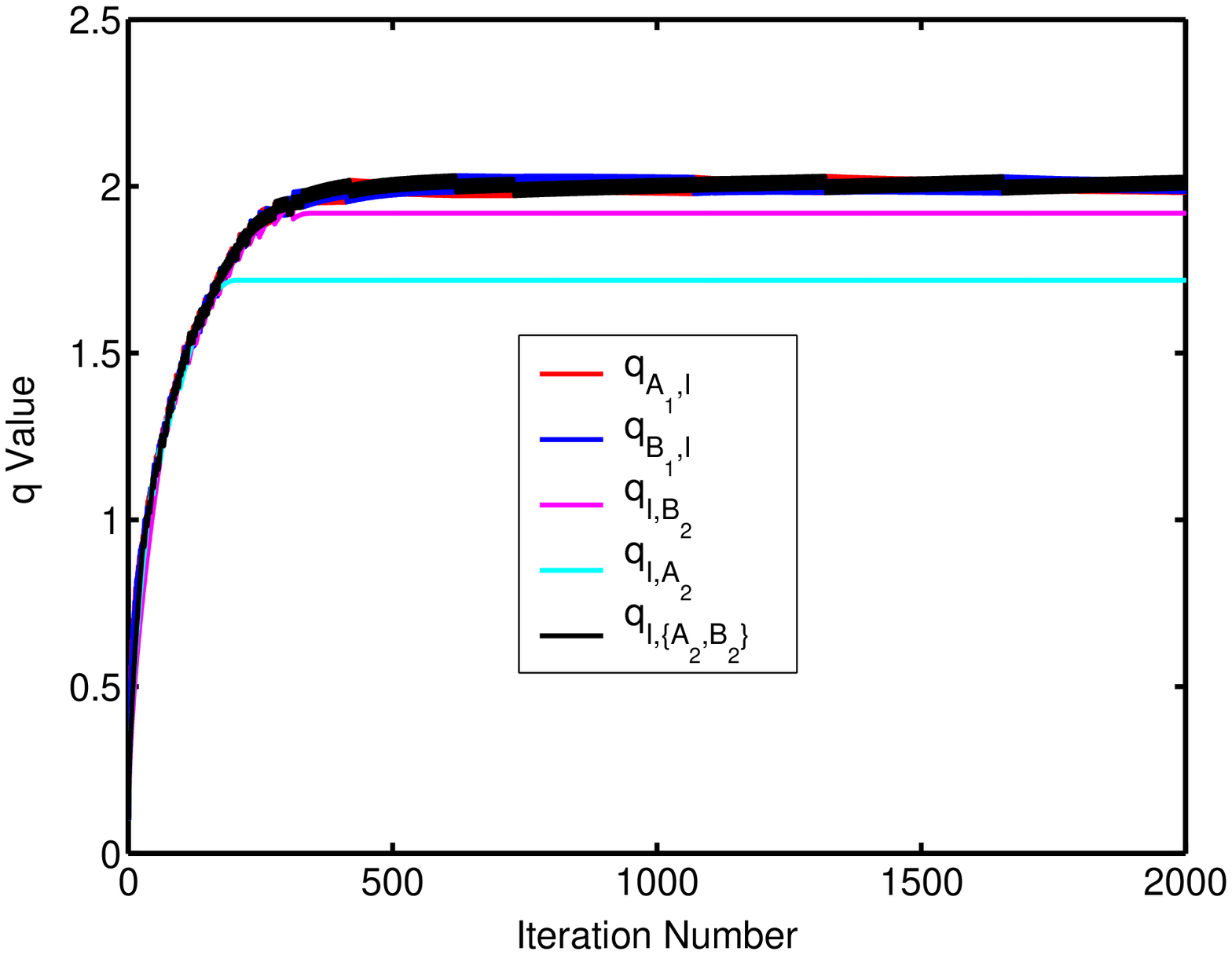}}}
\end{center}
\begin{center}
\caption{Convergence results for the X topology presented in Fig.~\ref{fig:x_topology}. The total achieved rate approaches to the optimum throughput 0.66. The optimum throughput is 0.50 when there is no network coding. $C_1=C_2=C_3=C_4=1$. } \label{fig:num_calc_cross_case1}
\end{center}
\vspace{5pt}
\end{figure*}

\begin{figure*}[t!]
\begin{center}
\subfigure[Rate]{{\includegraphics[width=5cm]{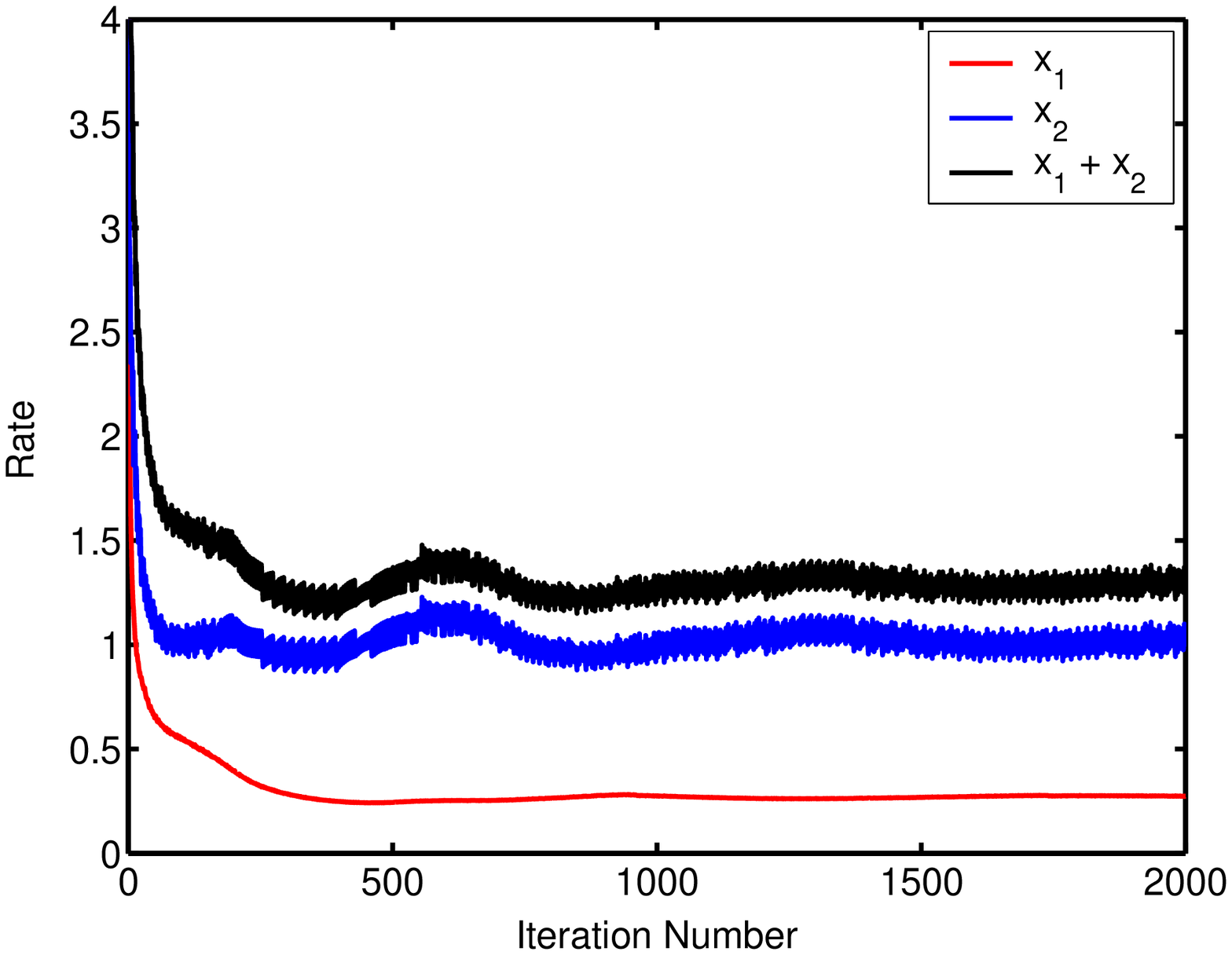}}}
\subfigure[Lagrange multipliers]{{\includegraphics[width=5cm]{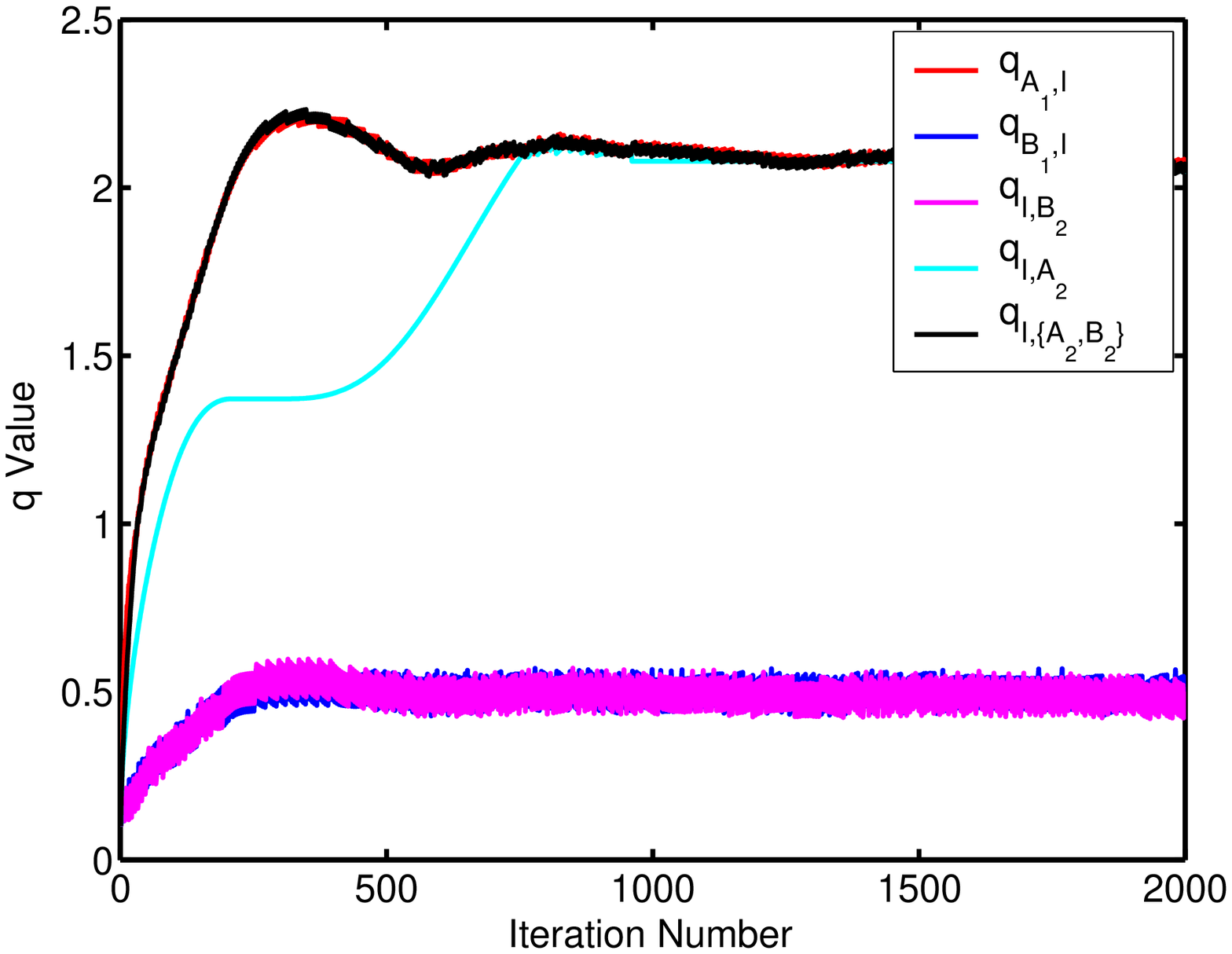}}}
\end{center}
\begin{center}
\caption{Convergence results for the X topology presented in Fig.~\ref{fig:x_topology}. The total achieved rate approaches the optimum throughput 1.3. The optimum throughput is 0.80 when there is no network coding. $C_1=C_4=1$, $C_2=C_3=4$. } \label{fig:num_calc_cross_case2}
\end{center}
\vspace{0pt}
\end{figure*}

\section*{\label{sec:numerical_results}Appendix A: Numerical Results}
In this section, we present numerical results that demonstrate the convergence of the solutions of NUM problems for one-hop and multi-hop network coding.

\vspace{-10pt}
\subsection{One-hop Network Coding}

First, we consider the Alice-and-Bob topology presented in Fig.~\ref{fig:topologies}(a). We consider two cases for wireless channel capacities: (i) $C_1=C_2=1$ units/transmission\footnote{We omit the units in the rest of the paper for brevity.}, and (ii) $C_1=1$, $C_2=4$. For the first case, the convergence of rates $x_1$, $x_2$, and $x_1 + x_2$ is presented in Fig.~\ref{fig:num_calc_A_B_case1}(a). One can see that the total rate $x_1 + x_2$ converges to $0.66$ which is the optimal achievable rate when network coding is used for this scenario. Note that total achieved throughput is $0.50$ for this scenario when network coding is not used. For the second case, it is seen in seen in Fig.~\ref{fig:num_calc_A_B_case2}(a) that the total throughput approaches to the optimal achievable rate of $0.88$ when network coding is used. Note that the total achievable throughput is $0.80$ when network coding is not used. We also present the convergence of Lagrange multipliers; $q_{A_1,I}$, $q_{A_2,I}$, $q_{I,A_2}$, $q_{I,A_1}$, and $q_{I,\{A_1,A_2\}}$ for both cases in Fig.~\ref{fig:num_calc_A_B_case1}(b) and Fig.~\ref{fig:num_calc_A_B_case2}(b), respectively.

Second, we consider the X topology presented in Fig.~\ref{fig:x_topology}. We consider two cases for wireless channel capacities: (i) $C_1=C_2=C_3=C_4=1$, and (ii) $C_1=C_4=1$, $C_2=C_3=4$. In both cases the total rate $x_1 + x_2$ approaches to the optimum achievable rates; $0.66$ and $1.3$ as seen in Fig.~\ref{fig:num_calc_cross_case1}(a) and Fig.~\ref{fig:num_calc_cross_case2}(a). We also show results for the convergence of the Lagrange multipliers for both cases in Fig.~\ref{fig:num_calc_cross_case1}(b) and Fig.~\ref{fig:num_calc_cross_case2}(b).

\subsection{Multi-Hop Network Coding}
\begin{figure*}[t!]
\begin{center}
\subfigure[Rate]{{\includegraphics[width=5cm]{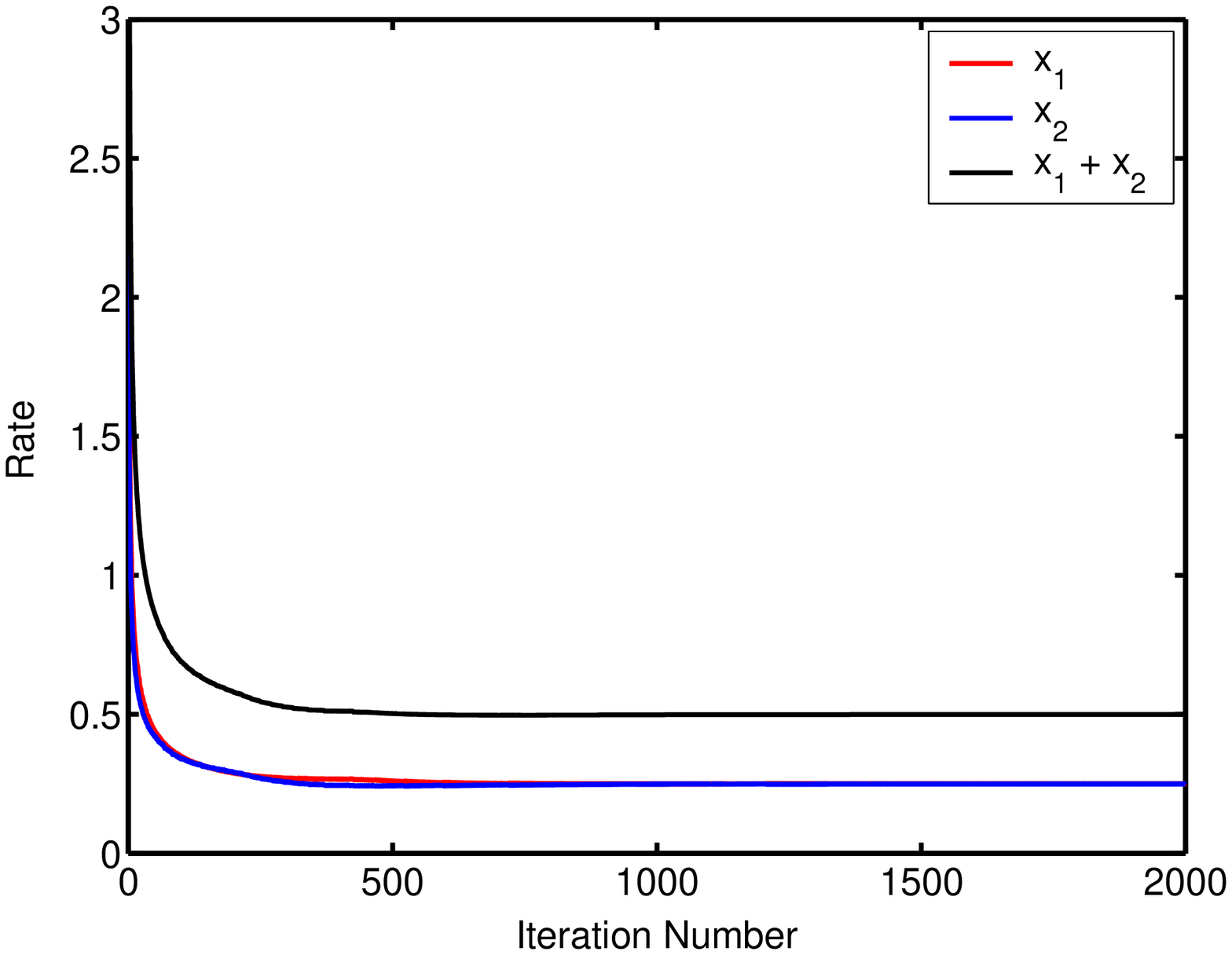}}}
\subfigure[Lagrange multipliers]{{\includegraphics[width=5cm]{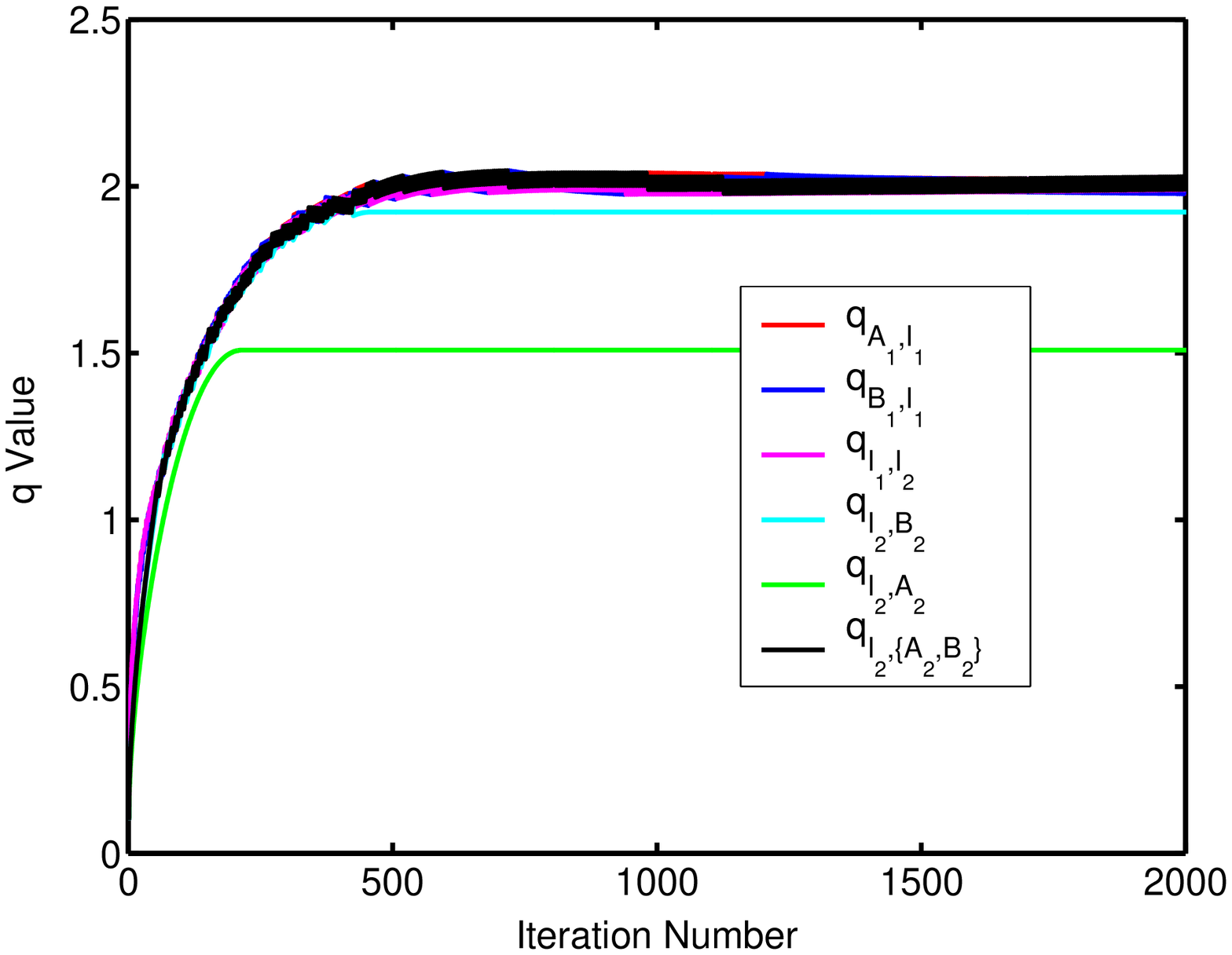}}}
\end{center}
\begin{center}
\caption{Convergence results for the butterfly topology presented in Fig.~\ref{fig:butterfly_topology}. The total achieved rate approaches the optimum throughput 0.50. The optimum throughput is 0.33 when there is no network coding. $C_1=C_2=C_3=C_4=C_5=1$. } \label{fig:num_calc_butterfly_case1}
\end{center}
\end{figure*}

\begin{figure*}[t!]
\begin{center}
\subfigure[Rate]{{\includegraphics[width=5cm]{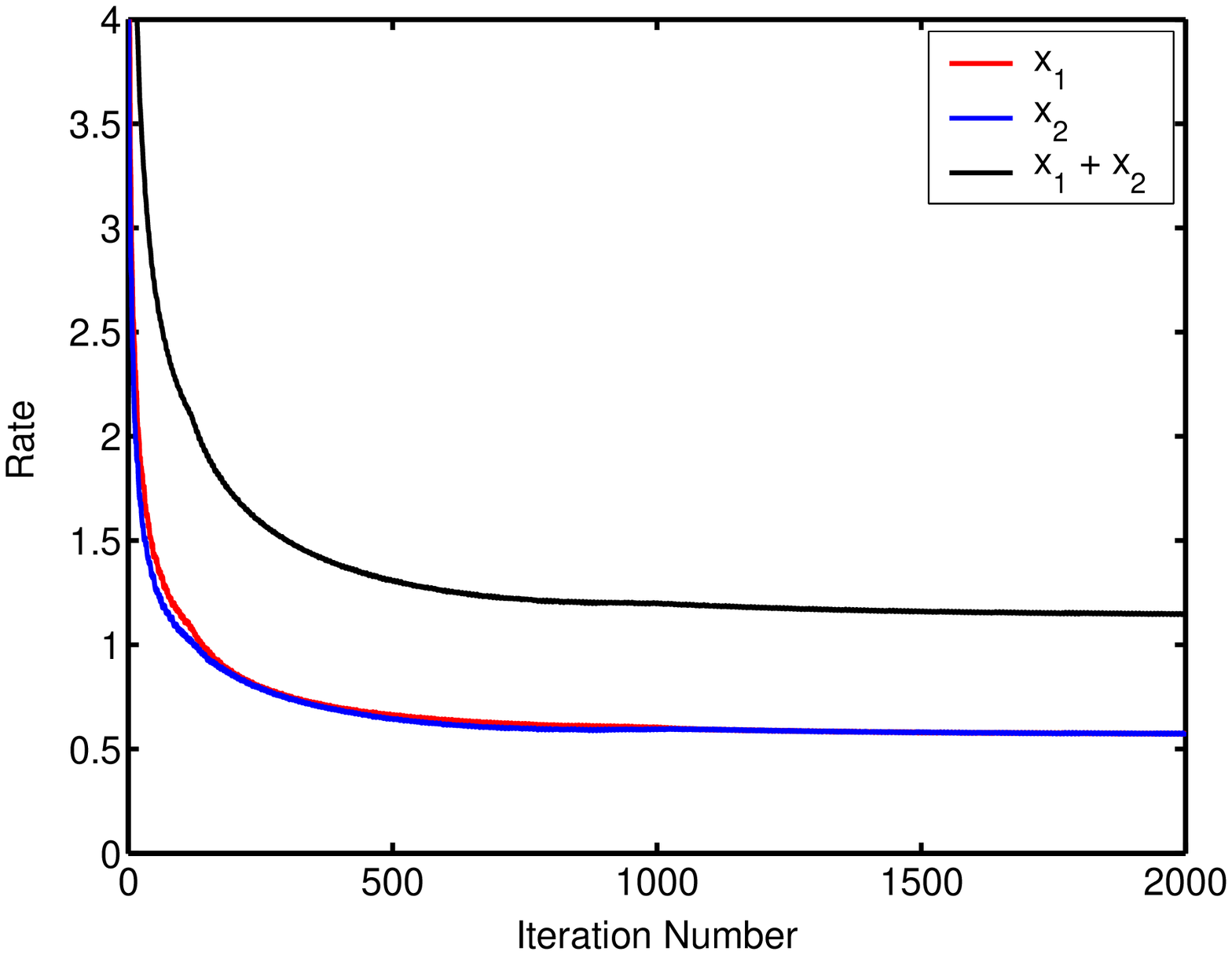}}}
\subfigure[Lagrange multipliers]{{\includegraphics[width=5cm]{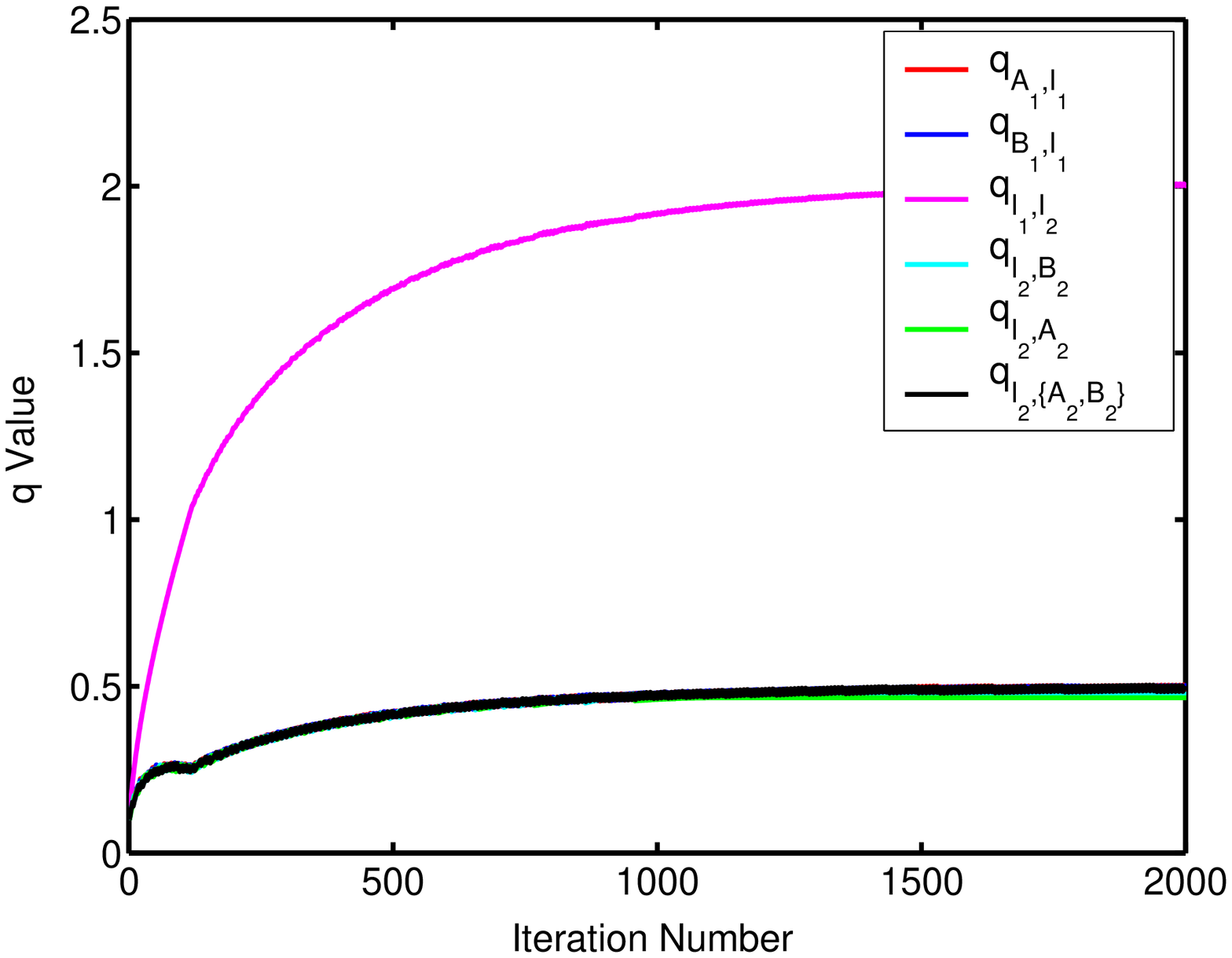}}}
\end{center}
\begin{center}
\caption{Convergence results for the butterfly topology presented in Fig.~\ref{fig:butterfly_topology}. The total achieved rate approaches the optimum throughput 1.14. The optimum throughput is 0.66 when there is no network coding. $C_1=C_2=C_4=C_5=4$, $C_3=1$.} \label{fig:num_calc_butterfly_case2}
\end{center}
\end{figure*}

We consider the butterfly topology presented in Fig.~\ref{fig:butterfly_topology}.
We consider two scenarios for the wireless channel capacities; (i) $C_1=C_2=C_3=C_4=C_5=1$ and (ii) $C_1=C_2=C_4=C_5=4$, $C_3=1$. The total rate approaches the optimal achievable rate in both scenarios: $0.5$ for the first case as shown in Fig.~\ref{fig:num_calc_butterfly_case1}(a), and $1.14$ for the second case as shown in Fig.~\ref{fig:num_calc_butterfly_case2}(a). In  both scenarios, we show the convergence of the  Lagrange multipliers, Fig.~\ref{fig:num_calc_butterfly_case1}(b) and Fig.~\ref{fig:num_calc_butterfly_case2}(b).

\end{document}